\documentclass[]{mn2e}
\usepackage{epsfig}
\def\gsim{\vcenter{\hbox{$>$}\offinterlineskip\hbox{$\sim$}}}
\def\lsim{\vcenter{\hbox{$<$}\offinterlineskip\hbox{$\sim$}}}
\title[Spectral atlas of $\omega$\,Centauri]{A spectral atlas of
post-main-sequence stars in $\omega$\,Centauri: kinematics, evolution,
enrichment and interstellar medium}
\author[Jacco Th. van Loon et al.]{Jacco Th. van Loon$^{1}$\thanks{E-mail:
jacco@astro.keele.ac.uk}, Floor van Leeuwen$^{2}$, Barry Smalley$^{1}$, Andrew
W. Smith$^{1}$,
\newauthor
Nicola A. Lyons$^{1}$, Iain McDonald$^{1}$, Martha L. Boyer$^{3}$\\
$^{1}$Astrophysics Group, Lennard-Jones Laboratories, Keele University,
      Staffordshire ST5 5BG, United Kingdom\\
$^{2}$Institute of Astronomy, Madingley Road, Cambridge CB3 0HA, United
      Kingdom\\
$^{3}$Department of Astronomy, 116 Church Street SE, University of Minnesota,
      Minneapolis, MN 55455, USA}
\date{Submitted July 2007}
\pagerange{\pageref{firstpage}--\pageref{lastpage}}
\pubyear{2007}
\begin{document}
\maketitle
\label{firstpage}
\begin{abstract}
We present a spectral atlas of the post-main-sequence population of the most
massive Galactic globular cluster, $\omega$\,Centauri. Spectra were obtained
of more than 1500 stars selected as uniformly as possible from across the (B,
B--V) colour-magnitude diagram of the proper motion cluster member candidates
of van Leeuwen et al.\ (2000). The spectra were obtained with the 2dF
multi-fibre spectrograph at the Anglo Australian Telescope, and cover the
approximate range $\lambda\sim3840$--4940 \AA\ at a resolving power of
$\lambda/\Delta\lambda\simeq2000$. This constitutes the most comprehensible
spectroscopic survey of a globular cluster. We measure the radial velocities,
effective temperatures, metallicities and surface gravities by fitting {\sc
atlas9} stellar atmosphere models. We analyse the cluster membership and
stellar kinematics, interstellar absorption in the Ca\,{\sc ii} K line at 3933
\AA, the RR Lyrae instability strip and the extreme horizontal branch, the
metallicity spread and bimodal CN abundance distribution of red giants,
nitrogen and s-process enrichment, carbon stars, pulsation-induced Balmer line
emission on the asymptotic giant branch (AGB), and the nature of the post-AGB
and UV-bright stars. Membership is confirmed for the vast majority of stars,
and the radial velocities clearly show the rotation of the cluster core. We
identify long-period RR\,Lyrae-type variables with low gravity, and
low-amplitude variables coinciding with warm RR\,Lyrae stars. A barium
enhancement in the coolest red giants indicates that $3^{\rm rd}$ dredge-up
operates in AGB stars in $\omega$\,Cen. This is distinguished from the
pre-enrichment by more massive AGB stars, which is also seen in our data. The
properties of the AGB, post-AGB and UV-bright stars suggest that RGB mass loss
may be less efficient at very low metallicity, [Fe/H]$\ll-1$, increasing the
importance of mass loss on the AGB. The catalogue and spectra are made
available via CDS.
\end{abstract}

\section{Introduction}

The most massive Galactic globular cluster, $\omega$\,Centauri (NGC\,5139) is
of considerable interest both for astrophysics and cosmology. Being the most
populous star cluster in our galaxy, one has a chance to catch stars in their
most rapid phases of post-main-sequence evolution, such as post-Asymptotic
Giant Branch (post-AGB) stars which evolve on timescales of $10^5$ years or
less to become white dwarfs after significant mass return into the
interstellar medium (ISM). Its proximity, $d\sim5$ kpc (e.g., Del Principe et
al.\ 2006; van de Ven et al.\ 2006; Caputo et al.\ 2002), and relatively low
foreground extinction, $E(B-V)\sim0.11$ mag (Lub 2002), further aid in
detailed studies of its constituents. Besides, $\omega$\,Cen may hold vital
clues to understanding the assembly of massive galaxies and the fate of the
many satellites predicted to have formed in the $\Lambda$-CDM cosmological
paradigm (Klypin et al.\ 1999). Understanding its dynamical history and
possible origin within a larger galactic body are key to understanding the
nature of ultra-massive globular clusters such as G\,1 in the Local Group
spiral galaxy M\,31 (e.g., Gebhardt et al.\ 2002), the ultra-compact dwarf
galaxies found in the Fornax cluster (Drinkwater et al.\ 2004), and the tidal
dissolution of stellar systems that then populate the Galactic halo with field
stars.

It has long been known that $\omega$\,Centauri hosts stars spanning a range of
metallicities, contrary to the paradigm that globular clusters are co-eval,
chemically uniform systems. Norris et al.\ (1996) determined calcium
abundances of bright Red Giant Branch (RGB) stars and found that about 80 per
cent belong to a metal-poor component now identified with a metallicity [Fe/H]
$\simeq-1.7$, with the remaining 20 per cent constituting a metal-richer
component, of [Fe/H] $\simeq-1.2$. Lee et al.\ (1999) and Pancino et al.\
(2000, 2002), using optical colour-magnitude diagrams, discovered a third
sub-population in the form of an anomalous feature in the RGB morphology, the
RGB-a; these red giants are, with [Fe/H] $\simeq-0.6$, an order of magnitude
more metallic than the main population of $\omega$\,Cen.

Lee et al. (1999) proposed an extended period of star formation, lasting
$\sim2$ Gyr, to explain the different populations through a chemical
enrichment scenario within a larger, now-disrupted system, a scenario proposed
previously also by Zinnecker et al.\ (1988) and Freeman (1993). Although
Stanford et al.\ (2006a) do indeed find that the metal-rich stars are the
youngest by 2-4 Gyr, Sollima et al.\ (2005) find that the metal-rich and
intermediate-metallicity populations must have formed within 2 Gyr from the
oldest metal-poor stars. The situation may be more complicated: Villanova et
al.\ (2007) suggest that the metal-rich stars and a third of the metal-poor
stars are the {\it oldest}, with the remaining two-thirds of metal-poor stars
and the intermediate-metallicity stars being younger by 3-4 Gyr.

A double main sequence was discovered by Anderson (1997; see also Bedin et
al.\ 2004), and interpreted by Norris (2004) in terms of the three RGB
sub-populations; counter-intuitively, the {\it less} metal-poor population
must correspond to the {\it bluer} main sequence (confirmed spectroscopically
by Piotto et al.\ 2005). Norris explained this by the intermediate-metallicity
population being enhanced in helium. Though reconciling the observations, the
helium enhancement poses a challenge for chemical evolution models, as neither
all of the cluster sub-giant stars nor all the intermediate-metallicity
RR\,Lyrae variables on the Horizontal Branch (HB) are helium-rich (Villanova
et al.\ 2007; Sollima et al.\ 2006).

The mixed stellar populations complicate studies of stellar evolution in
$\omega$\,Cen, but on the other hand it offers the possibility to compare the
evolution of stars between these different populations. Questions that remain
to be answered include the mechanism of mass loss on the RGB, how (and why) it
varies between different stars, and how this is reflected in the properties of
the HB (see Catelan 2005 for a review), in particular the extreme HB (D'Cruz
et al.\ 1996). A type of HB star, RR\,Lyrae-type pulsators are distance
indicators. The RR\,Lyrae-derived distance to $\omega$\,Cen ($d=5.5\pm0.04$
kpc: Del Principe et al.\ 2006) differs from that derived from the internal
kinematics ($d=4.8\pm0.3$ kpc: van de Ven et al.\ 2006), and it is therefore
important to reach a full understanding of the RR\,Lyrae phenomenon. Different
types of RR\,Lyrae exist in $\omega$\,Cen, pulsating in different modes, and
stars are found with colours and magnitudes that overlap with those of
RR\,Lyrae but which are not strong pulsators (Sandage \& Katem 1968). It is
not clear to what extent this is due to different evolutionary stages,
differences in RGB mass loss, or a result of the spread in elemental
abundances.

During the subsequent AGB and post-AGB evolution additional mass is returned
to the ISM, possibly enriched in s-process elements and/or dust grains. It is
not clear, though, how many globular cluster stars reach the AGB, how many of
those experience $3^{\rm rd}$ dredge-up, and whether globular cluster carbon
stars must all have formed through external pollution. It is also not
understood how AGB mass loss and dust formation proceeds at a considerably
sub-solar metallicity (van Loon 2006). A different question altogether is how
much post-AGB stars contribute to the UV light from old stellar systems such
as elliptical galaxies; this depends on the zero-age post-AGB luminosity
function and the timescale of post-AGB evolution.

The present study was motivated by the van Leeuwen et al.\ (2000) survey, who
determined the proper motion membership probabilities of nearly 10,000 stars
within half a degree from the centre of the cluster down to a photographic
magnitude of $\sim16.5$. We obtained medium-resolution follow-up spectroscopy
in the B-band of more than 1500 probable members. Whilst other follow-up
studies concentrated on obtaining radial velocities to map the 3-dimensional
kinematical structure of the cluster (Reijns et al.\ 2006) our survey covers
sufficient line diagnostics to be able to also automatically determine
temperatures, gravities and metallicities through fitting of the synthetic
spectra of model atmospheres. This allows us to break the degeneracies
encountered in colour-magnitude diagrams, and to study the effect of
metallicity on stellar evolution. Abundances of carbon, nitrogen and s-process
elements such as barium can be compared to assess the degree of enrichment
from internal nucleosynthesis and external pollution. The survey also provides
candidates for more detailed follow-up studies, including the use of blue-HB
stars as torches shining through the intervening ISM. Preliminary results were
presented in van Loon (2002), and the spectra are currently used to support
our {\it Spitzer Space Telescope} IRAC+MIPS atlas of the cluster (Boyer et
al.\ 2007; McDonald et al., in preparation). The remainder of this paper is
organised as follows:\\

\noindent
{\bf 2} Spectroscopy with the AAT/2dF\\
\hspace*{3mm}{\bf 2.1} Observations\\
\hspace*{3mm}{\bf 2.2} Data reduction\\
{\bf 3} The target sample\\
{\bf 4} The database\\
\hspace*{3mm}{\bf 4.1} The measurement of stellar parameters\\
\hspace*{3mm}{\bf 4.2} Data quality and reliability of the model fits\\
\hspace*{3mm}{\bf 4.3} Description of the electronic database\\
{\bf 5} Results\\
\hspace*{3mm}{\bf 5.1} Cluster membership and internal kinematics\\
\hspace*{3mm}{\bf 5.2} Interstellar absorption by ionised gas\\
\hspace*{3mm}{\bf 5.3} Metallicity and the morphology of the Hertzsprung-
\hspace*{9mm}Russell Diagram\\
\hspace*{3mm}{\bf 5.4} The first ascent Red Giant Branch (RGB)\\
\hspace*{9mm}{\bf 5.4.1} The nature of the anomalous RGB branch\\
\hspace*{9mm}{\bf 5.4.2} Nitrogen enrichment\\
\hspace*{9mm}{\bf 5.4.3} Chromospheric activity\\
\hspace*{3mm}{\bf 5.5} The RR\,Lyrae instability strip\\
\hspace*{3mm}{\bf 5.6} Extreme Horizontal Branch stars (EHB)\\
\hspace*{3mm}{\bf 5.7} The Asymptotic Giant Branch (AGB)\\
\hspace*{9mm}{\bf 5.7.1} M-type stars\\
\hspace*{9mm}{\bf 5.7.2} Pulsation\\
\hspace*{9mm}{\bf 5.7.3} Dredge-up\\
\hspace*{3mm}{\bf 5.8} Carbon stars\\
\hspace*{3mm}{\bf 5.9} Post-AGB and UV-bright stars\\
{\bf 6} Discussion\\
\hspace*{3mm}{\bf 6.1} Late stages of evolution of metal-poor stars\\
\hspace*{9mm}{\bf 6.1.1} Horizontal Branch stars and RR\,Lyrae variables\\
\hspace*{9mm}{\bf 6.1.2} Asymptotic Giant Branch stars and dredge-up\\
\hspace*{9mm}{\bf 6.1.3} Mass loss and post-giant branch evolution\\
\hspace*{3mm}{\bf 6.2} Cluster formation and evolution\\
\hspace*{9mm}{\bf 6.2.1} The nature of the multiple stellar populations\\
\hspace*{9mm}{\bf 6.2.1} Gas retention and accretion\\
{\bf 7} Summary of conclusions

\section{Spectroscopy with the AAT/2dF}

\subsection{Observations}

The 2-degree Field (2dF) multi-fibre spectrograph at the Anglo Australian
Telescope (AAT) was used on the second half of the nights on 26 and 27
February and 1 and 2 March, 2000, in combination with the 1200B grating to
obtain spectra in the approximate range $\lambda\sim3840$--4940 \AA\ at a
resolving power of $\lambda/\Delta\lambda\simeq2000$. A log of the
observations is presented in Table 1.

%
%
\begin{table*}
\caption[]{Log of observations, with date and airmass given for the middle of
the exposures. The value for the seeing is only an indicative figure for the
conditions at the time of observation. The mean velocity, $\langle v_{\rm LSR}
\rangle$ was computed relative to the Local Standard of Rest (LSR) along with
the standard deviation, for the $N_{\rm cluster}$ probable cluster members ---
i.e.\ excluding $N_{\rm field}$ obvious foreground stars.}
\begin{tabular}{lcccccrrrrll}
\hline\hline
ID                   &
UT date              &
airmass              &
seeing               &
plate                &
CCD                  &
$t_{\rm exp}$ (s)    &
$N_{\rm cluster}$    &
$N_{\rm field}$      &
$N_{\rm sky}$        &
selection            &
$\langle v_{\rm LSR} \rangle$ \\
\hline
1a                   &
2000 02 26.72        &
1.04                 &
$1.3^{\prime\prime}$ &
1                    &
1                    &
$4\times300$         &
41                   &
1                    &
120                  &
$B<13.5$             &
$234\pm16$           \\
1b                   &
...                  &
...                  &
...                  &
...                  &
2                    &
...                  &
44                   &
-                    &
116                  &
...                  &
$232\pm14$           \\
2a                   &
2000 02 26.79        &
1.11                 &
$1.7^{\prime\prime}$ &
0                    &
1                    &
$3\times300$         &
14                   &
1                    &
143                  &
...                  &
$243\pm13$           \\
2b                   &
...                  &
...                  &
...                  &
...                  &
2                    &
...                  &
33                   &
-                    &
131                  &
...                  &
$232\pm13$           \\
3a                   &
2000 02 27.64        &
1.11                 &
$2.0^{\prime\prime}$ &
0                    &
1                    &
$3\times1200$        &
134                  &
5                    &
31                   &
$B\geq 13.5$         &
$247\pm15$           \\
3b                   &
...                  &
...                  &
...                  &
...                  &
2                    &
...                  &
124                  &
2                    &
36                   &
...                  &
$227\pm14$           \\
4a                   &
2000 02 27.73        &
1.04                 &
$2.0^{\prime\prime}$ &
1                    &
1                    &
$3\times1500$        &
126                  &
3                    &
35                   &
...                  &
$241\pm18$           \\
4b                   &
...                  &
...                  &
...                  &
...                  &
2                    &
...                  &
126                  &
4                    &
38                   &
...                  &
$232\pm14$           \\
5a                   &
2000 03 01.55        &
1.43                 &
$1.5^{\prime\prime}$ &
1                    &
1                    &
$3\times1500$        &
121                  &
1                    &
47                   &
...                  &
$246\pm15$           \\
5b                   &
...                  &
...                  &
...                  &
...                  &
2                    &
...                  &
128                  &
-                    &
28                   &
...                  &
$238\pm14$           \\
6a                   &
2000 03 01.64        &
1.11                 &
$1.2^{\prime\prime}$ &
0                    &
1                    &
$3\times1500$        &
111                  &
1                    &
41                   &
...                  &
$244\pm14$           \\
6b                   &
...                  &
...                  &
...                  &
...                  &
2                    &
...                  &
117                  &
1                    &
47                   &
...                  &
$240\pm12$           \\
7a                   &
2000 03 01.74        &
1.06                 &
$2.2^{\prime\prime}$ &
1                    &
1                    &
$3\times1500$        &
109                  &
-                    &
48                   &
...                  &
$240\pm17$           \\
7b                   &
...                  &
...                  &
...                  &
...                  &
2                    &
...                  &
103                  &
-                    &
43                   &
...                  &
$234\pm11$           \\
8a                   &
2000 03 02.64        &
1.10                 &
$1.4^{\prime\prime}$ &
0                    &
1                    &
$3\times1500$        &
100                  &
1                    &
53                   &
...                  &
$243\pm16$           \\
8b                   &
...                  &
...                  &
...                  &
...                  &
2                    &
...                  &
105                  &
2                    &
47                   &
...                  &
$232\pm11$           \\
9a                   &
2000 03 02.73        &
1.05                 &
$2.5^{\prime\prime}$ &
1                    &
1                    &
$4\times1500$        &
88                   &
4                    &
72                   &
...                  &
$246\pm16$           \\
9b                   &
...                  &
...                  &
...                  &
...                  &
2                    &
...                  &
103                  &
1                    &
59                   &
...                  &
$235\pm11$           \\
\hline
\end{tabular}
\end{table*}

Each of the 400 fibres has a diameter of $2^{\prime\prime}$ projected on the
sky, limiting spectroscopy in crowded stellar fields. The proximity of
$\omega$\,Cen, the fact that it has a relatively low stellar density, and a
careful selection procedure (Section 3) enables us to obtain uncontaminated
spectra of the majority of post-main-sequence stars in this cluster. Despite
its large angular size, $\omega$\,Cen does not fill the full $2^\circ$ field,
and fibre allocation was therefore limited by the physical size of the fibre
buttons and crossings of the fibre arms. The field was acquired using four
fiducial stars. These were selected from stars with a high cluster membership
probability (Section 3) to safeguard the relative astrometry of the targets.
Subsequent fields were observed using alternating fibre-positioning plates.

Because spectra are displayed adjacently on each of two Charge Coupled Devices
(CCDs), bright targets defined as having B-band magnitudes $B<13.5$ mag were
observed separately from fainter targets. The bright targets received three
exposures of 5 minutes each, whilst the individual exposures of the fainter
targets were normally 25 minutes in duration. Typically, in each observation
$\sim200$ fibres were used to obtain spectra of the sky; these positions were
distributed on a grid with a density declining with distance from the centre,
and they were checked for the absence of contaminating stellar light.

Zero-integration CCD readouts (``bias'' frames) were taken to measure the
electronic offset and readout noise properties, and long integrations with the
shutter closed were taken to measure the dark current. Exposures of an Fe/Ar
arc lamp spectrum were taken for wavelength calibration, at each telescope
pointing to alleviate the effects of flexure within the instrument. Exposures
of a quartz ``white-light'' lamp were also obtained, through the fibres at
each telescope pointing, to measure the pixel-to-pixel variations in the
response of the CCD. For each observation, three exposures of 5 minutes each
were obtained at a position off-set from the target field by $0.5^\circ$, to
measure the relative throughput of the fibres.

The observing conditions were typical, with the seeing varying between 1.2 and
$2.5^{\prime\prime}$, and some scattered clouds on the February nights. The
overall quality of the spectra was excellent, and the analysis was almost
never limited by the signal-to-noise (S/N) ratio even for the faintest targets
($B\sim16.5$ mag). S/N ratios computed from pixel-to-pixel variations between
repeat observations are typically $\sim50$ per pixel around $B=16$ mag to
$>$100 per pixel for $B<15$ mag in the redder half of the spectrum, and
$\sim30$ in the bluer half of the spectrum. The S/N ratio per spectral
resolution element is roughly a factor $\sqrt{2}$ higher than this.

\subsection{Data reduction}

The data were reduced using version 2.3 (June 2002) of 2dFdr. The positions of
the (curved) spectra on the CCD are parameterised in a default tram map, which
was fitted to the data by applying rotation and translation corrections. The
fits and fibre identifications were checked manually, updating the table with
broken fibres where necessary.

First the fibre flatfield and arc exposures were reduced. Then the off-set sky
frames were reduced, which included subtracting scattered light and
combination of the individual frames via their median (on a pixel-to-pixel
basis). The target fields were divided through by the flatfield, and scattered
light was subtracted. The spectra were extracted from the frames by fitting
the profile perpendicular to the direction of dispersion with a gaussian,
rejecting anomalous pixel values if they deviated by more than $20\times$ the
noise level. The multiple exposures of the same star on a given night were
combined via their mean, after adjusting the continuum levels (applying a
smoothing of 101 pixels) and rejecting pixel values that deviated by more than
five times the noise level.

Due in part to contaminating light in the off-set sky observations, the fibre
throughputs were ill-determined. This resulted in large errors in the sky
subtraction, in particular for the faintest targets. We therefore performed a
second-order sky subtraction: for each wavelength, up to 5 sky values at
either side (on the CCD) of the object fibre were averaged, excluding the
highest and lowest values. This also improved the correction for scattered
light on the CCD.

\section{The target sample}

The basis for the spectroscopic target sample was the proper motion survey by
van Leeuwen et al.\ (2000), which comprises 9847 stars, and we shall refer to
their catalogue numbers for star identifications (Leiden Identifier, or LEID).
Stars were selected for possible spectroscopic observation if they had a
cluster membership probability of at least 90 per cent (but see below for
variable stars), and both B and V-band photometry. A further 172 stars were
rejected because of neighbouring stars within $2^{\prime\prime}$ that could
contaminate the light within the fibre.

Next, a grid was overlaid on the (B, B--V) diagram, with a grid spacing of
$\Delta B=0.1$ and $\Delta(B-V)=0.05$. In each pass, the star within a grid
element that had the (next) largest distance from the cluster centre was
assigned the (next) highest priority. This alleviates the natural bias towards
the most populous parts of the cluster and colour-magnitude diagram. As a
result, our spectral atlas samples not only the central parts of the cluster
and the ``spine'' of the RGB and HB, but includes the cluster fringes (which
also makes fibre allocation easier) and rare objects in trace populations or
brief phases of evolution. In seven such passes, 2477 stars were picked, of
which 688 in the highest category. Of the remaining stars, 98 fiducial stars
were selected with $B<14$ mag (all of these have $(B-V)>1$ mag) and no
neighbours within $3^{\prime\prime}$. This left 4681 stars to be available for
fibre allocation at the lowest priority.

A further 15 stars were observed that did not satisfy the 90 per cent proper
motion membership probability criterion, but which are of special interest.
These comprise variable stars (mostly of RR\,Lyrae type) and the two brightest
(in the B-band) suspected members: post-AGB star candidates \#16018
(Fehrenbach's star; ROA\,24, Woolley 1966) and the long-period Cepheid
variable \#32029 (V1; $P=32$ d). To ensure sufficient coverage of the
RR\,Lyrae instability strip, the highest priority was assigned to all known
RR\,Lyrae stars and ``non-variable'' stars within the RR\,Lyrae ``box'', which
we defined as $0.3<(B-V)<0.6$ and $14.5<B<15.2$. Other stars of special
interest include the longest-period confirmed variable star member of
$\omega$\,Cen, \#44262 (V42; $P=149$ d), and two known carbon stars, \#41071
(ROA\,577) and \#52030 (ROA\,55). The long-period variable star \#34041
($P=508$ d) was not observed, as its proper motion clearly places it outside
of the cluster. In total, 1766 spectra were obtained, including 142 RR\,Lyrae
and 238 duplicate observations of the same star.

%
%
\begin{figure}
\centerline{\psfig{figure=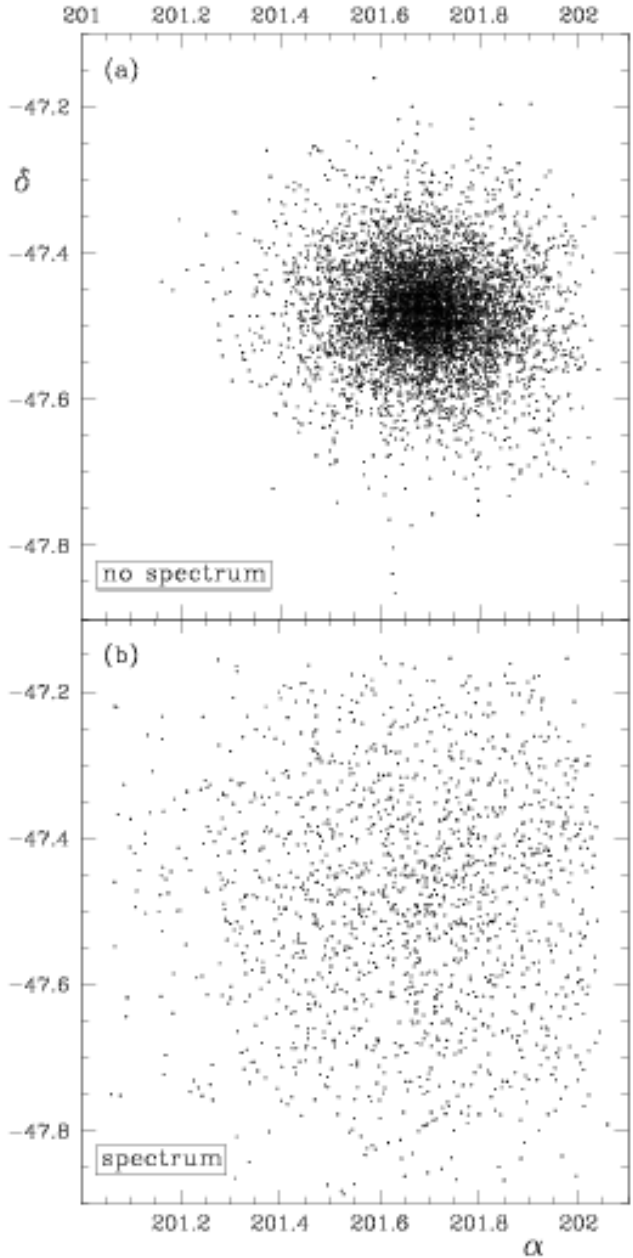,width=84mm}}
\caption[]{Distribution on the sky of (a) all the potential targets minus the
observed stars and (b) the stars that were actually observed. There is very
little bias towards the centre of the cluster, without avoiding it altogether,
and the outskirts are especially well sampled.}
\end{figure}

%
%
\begin{figure}
\centerline{\psfig{figure=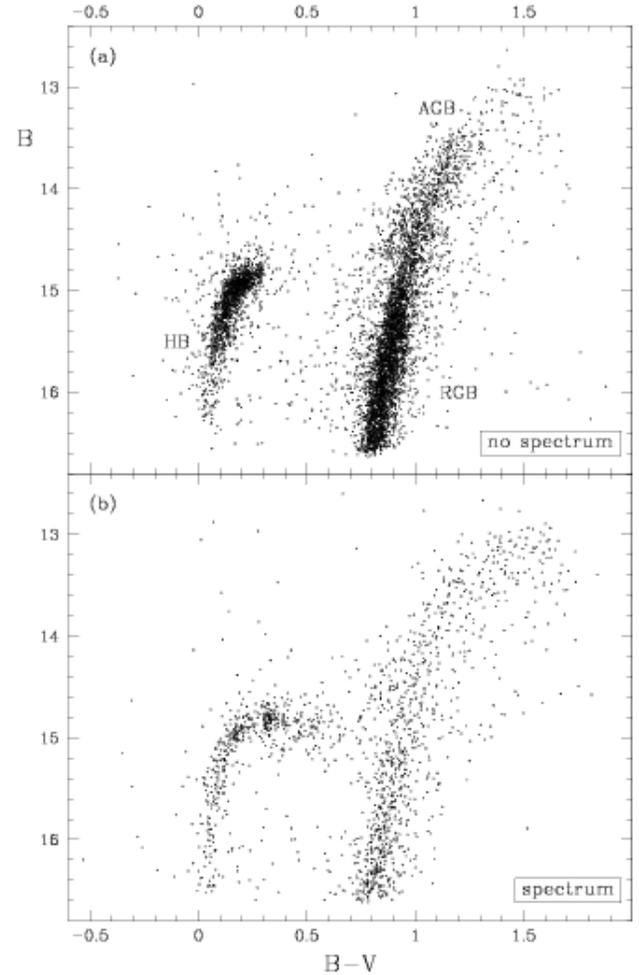,width=84mm}}
\caption[]{Distribution in the (B, B--V) diagram of (a) all the potential
targets minus the observed stars and (b) the stars that were actually
observed. All regions in the diagram all sampled, with the only strong bias
remaining being the RR\,Lyrae variables and non-variables with similar colours
and magnitudes.}
\end{figure}

The resulting distribution on the sky of the observed stars shows that all of
the cluster is well sampled, from the outskirts into the core (Fig.\ 1).
Similarly, the observations sample the colour-magnitude diagram very well
(Fig.\ 2), with a greatly reduced bias towards the densest populated parts of
the giant branches. They cover the colour spread at the top of the RGB and
AGB, the anomalous branch at $(B-V)>1$ and $B\sim15$, the extreme blue HB
(which turns vertical in a (B, B--V) diagram), and several ``UV-bright'' stars
above and to the blue of the HB.

\section{The database}

\subsection{The measurement of stellar parameters}

Each 2dF spectrum was matched with a model spectrum in an automatic fashion.
To this end, 2503 {\sc atlas9} model atmospheres (Kurucz 1993) were generated
on a grid of stellar parameters, viz.\ effective temperature, $T_{\rm eff}$,
metallicity (scaled solar abundances), [Fe/H], and surface effective gravity,
$\log g_{\rm eff}$ (Fig.\ 3). Synthetic spectra were computed for these models
with {\sc synthe} (Kurucz 1993), at the spectral resolving power of the 2dF
spectra but sampling at a ten times higher rate to enable measurement of the
radial velocity shift to a precision of a few km s$^{-1}$ --- even though the
{\it accuracy} of this value is less (see Sect.\ 4.2). The spectra were
interpolated to increase the grid resolution to steps of $\Delta$[Fe/H]=0.25.

%
%
\begin{figure}
\centerline{\psfig{figure=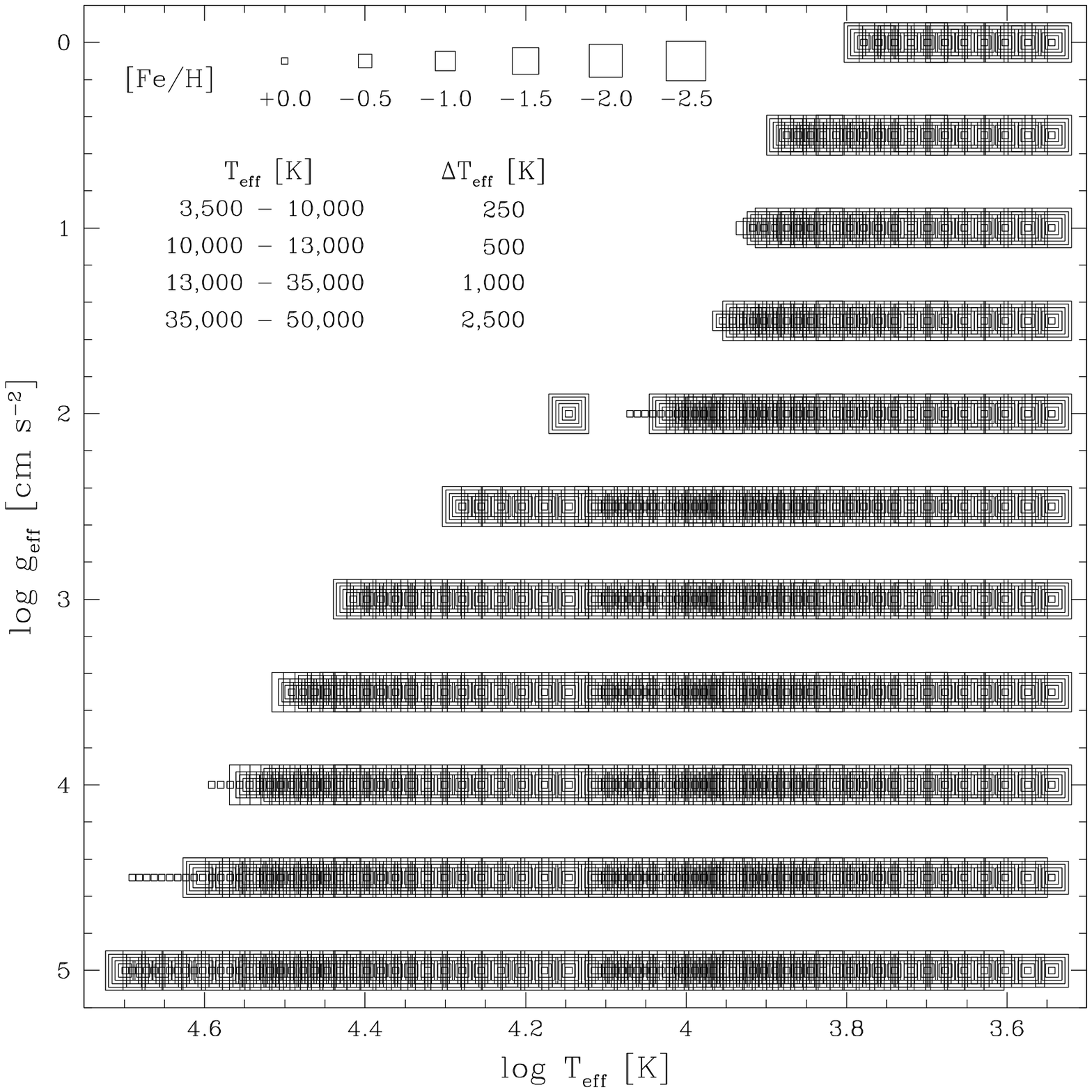,width=84mm}}
\caption[]{Grid of {\sc atlas9} model atmospheres that were generated and for
which synthetic spectra were computed with {\sc synthe}. Spectra were
interpolated to provide approximate spectra for metallicities of
[Fe/H]=$-0.25$, $-0.75$, $-1.25$, $-1.75$ and $-2.25$.}
\end{figure}

The atomic line list used in the models is quite extensive and includes many
(but not all) transitions in the CH, CN and TiO molecules (Kurucz 1999).
Besides omissions such as the C$_2$ molecule which is seen in carbon stars
(Sect.\ 5.8), some of the atomic data are uncertain (e.g., helium, Sect.\ 5.6)
or the atmosphere model has difficulty reproducing the exact strength (as in
the case of the molecules). Abundance anomalies are not uncommon in
post-main-sequence stars as a result of surface enrichment with specific
products of nucleosynthesis, most spectacularly seen in carbon stars, and
$\omega$\,Cen is known for the abundance anomalies seen even in its
main-sequence stars (Stanford et al.\ 2006b, 2007). Nevertheless, these models
produce useful stellar classifications for a global analysis of the cluster
population and stellar evolution, and as a starting point for more detailed
studies of individual (classes of) objects.

Expanding the observed spectrum, $f$, in each spectral point $a$ as a Taylor
series to second order,
\begin{equation}
f(x) \simeq f(a) + (x-a) \left. \frac{{\rm d}f}{{\rm d}x} \right|_{x=a} +
\frac{(x-a)^2}{2!} \left. \frac{{\rm d}^2f}{{\rm d}x^2} \right|_{x=a},
\end{equation}
and similarly for the model spectrum, $g$, we do not simply require that the
spectra reach similar values in each point, $f(a)\simeq g(a)$, but that the
{\it shape} of the spectra are similar in each point, $f(x)\simeq g(x)$. Thus,
with $N$ spectral points and $\Delta_a=f(a)-g(a)$, we minimise the statistic
\begin{equation}
\chi^2 = \sum_{i=1}^{N} \left( \frac{\Delta_{i+2} +\Delta_{i+1} -\Delta_{i-1}
+\Delta_{i-2}}{2} \right)^2
\end{equation}
(see Appendix A). It tests for similarity in intensity, slope, and curvature
at any given point in the spectrum. It measures correlated behaviour of the
residuals and thus enhances the ability to detect and correctly reproduce weak
spectral features, in contrast to the uncorrelated behaviour of noise or
spectral mismatch.

Before evaluating $\chi^2$ the observed spectrum is first normalised to the
continuum in the model spectrum, using a running boxcar with a width of 100
pixels ($\sim100$ \AA). The result of this procedure is that the fitting is
less sensitive to low-frequency structure in the spectrum, and thus less
affected by uncertainties in the wavelength-dependent instrumental throughput
or the exact level of the continuum. The latter is important in very cool
stars, where the atomic and molecular absorption veils much of the continuum.

The procedure is repeated after applying a velocity offset to the model
spectra using the non-relativistic Doppler formula. Initial offsets are
between $\Delta v=-200$ and 400 km s$^{-1}$ in steps of 100 km s$^{-1}$.
Subsequently, these steps are diminished by a factor 5 and applied around the
velocity at which $\chi^2$ was thus far minimal, with a final velocity step of
1 km s$^{-1}$.

As a compromise between maximising multiplexity and maximising fidelity, the
$\chi^2$ was evaluated within $\lambda=3910$ to 4940 \AA. Most spectra extend
beyond that, but the data near the edges are generally of lower quality
because of spectrum extraction and correction difficulties near the rims of
the CCDs. Also, the diminished sensitivity of the CCDs at ultraviolet
wavelengths below $\lambda\sim3900$ \AA\ reduces the S/N ratio, a problem in
particular for the red giants. Nevertheless, there are spectral features in
the ultraviolet that remain useful to be compared with the model spectrum that
best fits the rest of the observed spectrum, such as higher order lines of the
hydrogen Balmer series in the hotter stars and a strong CN band in some of the
cooler stars.

The H$\beta$, H$\gamma$ and H$\delta$ line profiles were checked for signs of
possible emission: if the maximum difference between the observed spectrum and
the best model spectrum was in excess of both 20 per cent of the continuum and
three times the standard deviation determined outside of the line profiles,
then the model fitting procedure was repeated excluding those points and the
two points at either side of them. A flag was set to 1 in the catalogue. The
same was done for the Ca\,H+K lines, which sometimes exhibit line emission
arising from a chromosphere. If detected, the flag was set to 2 (or 3 if both
the hydrogen and calcium lines showed signs of possible emission).

Bad data were removed; a spectrum was not considered if more than half of the
data were rejected, in which case the flag was set to 9. Including a few
spectra that obviously had no signal to speak of, a total of only 12 spectra
were rejected from the analysis, leaving 1754 classified spectra.

As an indication of, respectively, the total flux collected on the CCD and the
recorded spectral slope, we define a magnitude and colour that we compute from
the observed spectrum. The magnitude is broadly similar to the B-band, and is
defined as
\begin{equation}
B^\prime \equiv 25 - 2.5 \log \left( \sum_{i=1}^{N} f(i) \right),
\end{equation}
while the colour is defined in relation to the blue and red halves within this
band:
\begin{equation}
B_1-B_2 \equiv -2.5 \log \left( \frac{\sum_{i=1}^{N/2} f(i)}{\sum_{j=N/2}^{N}
f(j)} \right).
\end{equation}

\subsection{Data quality and reliability of the model fits}

Several checks can be made to ascertain the quality of the observed spectra
and the reliability of the model fits. In the following we will discuss the
accuracy of the radial velocities, the level of variation between repeat
observations, a comparison between the photometry and spectroscopic
measurements, and a number of spectral oddities that are not well reproduced
by the models.

The number statistics in Table 1 of the 1754 classified spectra includes a
computation of the mean velocity per CCD per observation. Without exception,
the mean velocity on CCD\,1 is larger than on CCD\,2: $\langle v_{\rm LSR}
\rangle_1 = 243\pm4$ km s$^{-1}$ versus $\langle v_{\rm LSR} \rangle_2 =
234\pm4$ km s$^{-1}$, where the errors are standard deviations. Because this
effect is larger than the variations between separate observations, we correct
for it by lowering all radial velocities from CCD\,1 by 4 km s$^{-1}$ and by
increasing all radial velocities from CCD\,2 by an equal amount.

%
%
\begin{table}
\caption[]{Average differences between repeat measurements.}
\begin{tabular}{lccccc}
\hline\hline
Selection                            &
$N$                                  &
$\Delta v$                           &
$\frac{\Delta T}{\langle T \rangle}$ &
$\Delta\log g$                       &
$\Delta$[Fe/H\rlap{]}                \\
                            &
                            &
\llap{(}km s$^{-1}$\rlap{)} &
                            &
\llap{(}cm s$^{-2}$\rlap{)} &
(dex)                       \\
\hline
All                      &
\llap{2}35               &
\llap{1}1.1              &
0.06                     &
0.90                     &
0.40                     \\
$B$--$V<0.3$             &
40                       &
\llap{1}3.4              &
0.13                     &
0.77                     &
0.62                     \\
$0.3\leq B$--$V\leq 0.6$ &
86                       &
\llap{1}0.8              &
0.04                     &
1.15                     &
0.27                     \\
$0.6<B$--$V<1$           &
82                       &
\llap{1}0.8              &
0.05                     &
0.85                     &
0.47                     \\
$B$--$V\geq 1$           &
27                       &
9.1                      &
0.04                     &
0.43                     &
0.31                     \\
\hline
\end{tabular}
\end{table}

Of the 238 repeat observations, 235 pairs of spectra were classified. The
average absolute values of the differences in the obtained parameters (Table
2) give us a rough estimate of the accuracy of these measurements, as the
variance of two measurements is half the square of their difference (Bland \&
Altman 1996). The velocity is thus determined with an accuracy of
$\sigma_v\sim11/\sqrt{2}=8$ km s$^{-1}$; slightly worse for the bluest stars
but better for the reddest --- this is almost certainly due to the increased
multiplexity of the cool star spectra. A similar effect is seen in the
relative error on the temperature, where the temperature is generally
determined with an accuracy of $\sim3$ to 4 per cent except for the bluest
stars ($B-V<0.3$) where the temperature cannot be determined to better than 9
per cent. The accuracy of the gravity seems better for the coolest stars, but
this may be misleading as some stars might have $\log(g)<0$ but we did not
compute models for such low gravities. The significantly increased spread in
$\log(g)$ values in the RR\,Lyrae box ($0.3\leq B-V\leq 0.6$) is probably due
to radial pulsation (see Section 5.5). The metallicity is determined with an
accuracy of $\sim0.2$ to 0.4 dex; our choice of steps in metallicity of 0.25
dex is therefore adequate.

%
%
\begin{figure}
\centerline{\psfig{figure=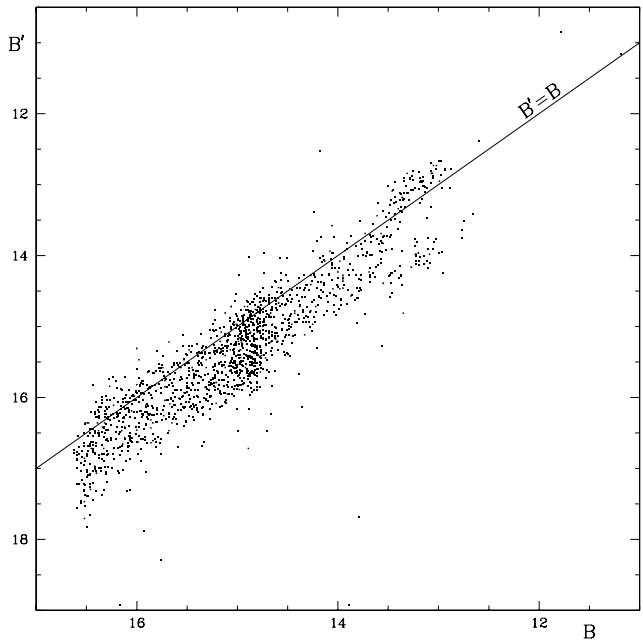,width=84mm}}
\caption[]{Comparison between the computed magnitude, $B^\prime$, and the
photometric magnitude, $B$. The absolute calibration of $B^\prime$ is rather
arbitrary, but there is a good correlation, with the spread of $\sim1$ mag due
to variations in the light losses in the fibres as a result of varying
observing conditions.}
\end{figure}

The quality of the spectra depends on the amount of light that enters the
fibre, which is measured by means of the computed magnitude $B^\prime$. After
normalising to an exposure time of 1000 seconds, it correlates well with the
photometric magnitude, $B$ (Fig.\ 4). The scatter of about a magnitude is
largely due to varying observing conditions such as seeing, transparency and
airmass, rather than inaccurate centering of the stars.  A few stars deviate
considerably more from the correlation; some of these are foreground stars, of
which the large proper motion difference with respect to the fiducial cluster
member stars is responsible for their poor centering and consequent light
losses. There is no sign of degradation in the ability to detect signal from
the faintest targets.

%
%
\begin{figure}
\centerline{\psfig{figure=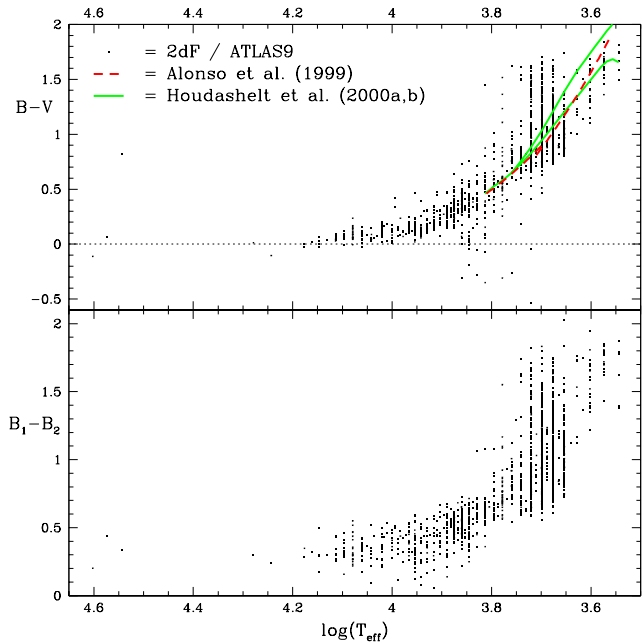,width=84mm}}
\caption[]{Comparison between the measured temperature, $T_{\rm eff}$, and the
photometric colour, $B-V$ (top panel) and the computed colour, $B_1-B_2$
(bottom panel). There is a good correlation in both cases; although the
scatter in blue $B_1-B_2$ is larger than in blue $B-V$, the $B_1-B_2$ does not
suffer from the large deviations seen in $B-V$ around $\log(T_{\rm
eff})\sim3.8$. In the top panel we also plot empirical relations for red giant
stars from Alonso et al.\ (1999) (dashed) and MARCS-based relations from
Houdashelt et al.\ (2000a,b) for $\log(g)=0$ (redder curve) and $\log(g)=1.5$
(bluer curve), all for [Fe/H]$=-1$ and reddened by $E(B-V)=0.11$ mag.}
\end{figure}

The computed colour, $B_1-B_2$ measures the spectral slope, and it is
therefore not surprising that it correlates well with the measured temperature
(Fig.\ 5). At $\log(T_{\rm eff})>3.9$ the scatter in the B$_1$--B$_2$ colours
is somewhat larger than in the B--V colours. On the other hand, the large
deviations in the B--V colours seen at $\log(T_{\rm eff})\sim3.8$ are not seen
in the B$_1$--B$_2$ colours --- none of these deviating data are from
RR\,Lyrae stars. The similarity between the two colours suggests that the
spectral slope is recorded correctly, and that the sky subtraction in the
bluest part of the spectrum is at least roughly correct.

Relationships between colour and temperature have often been used to derive
temperatures for stars from photometry. To compare some of these with our
results, we plot in the top panel of Fig.\ 5 the empirical relation from
Alonso et al.\ (1999) for red giant stars based on the infrared flux method
(dashed curve), and the relation from Houdashelt et al.\ (2000a,b) --- which
is based on MARC model atmospheres and empirical calibration --- where the
redder curve is for $\log(g)=0$ and the bluer curve is for $\log(g)=1.5$. We
selected the relations for [Fe/H]$=-1$, but the differences between the
relations covering the metallicity spread in $\omega$\,Cen are negligible. We
applied a reddening of $E(B-V)=0.11$ mag to these relations as the photometry
has not been dereddened. There is good agreement between these relations and
our results, in particular for the coolest stars. The redder stars at
$\log(T_{\rm eff})\sim3.8$ have lower gravity, which renders both our analysis
and the relations derived by Alonso et al.\ and Houdashelt et al.\ more
uncertain.

The models do not always reproduce all spectral features equally well. Apart
from the difficulty with molecular absorption, in particular in the case of
M-type and carbon stars, there are a few gravity-sensitive atomic lines which
can sometimes be clearly discrepant whilst the remainder of the spectrum is
reproduced accurately. Two cases were noted, where the discrepancy is more
than a factor two in equivalent width: (i) the Fe\,{\sc i} line at 4072 \AA\
is much weaker than predicted by the model, and (ii) the Ca\,{\sc i} line at
4226 \AA\ and the Fe\,{\sc i} lines at 4144 and 4384 \AA\ are much stronger
than predicted by the model. Care should be taken not to misinterpret these as
due to abundance anomalies. These lines are nevertheless included in the
fitting procedure as in many cases there is no discrepancy and they do not by
themselves have the ability to significantly alter the overall solution.

\subsection{Description of the electronic database}

The data products delivered to the community by our 2dF survey of
$\omega$\,Cen comprise of a catalogue of star identifiers with coordinates and
photometry and a number of quantities measured from the spectra and discussed
in this paper, and the full set of reduced spectra and model fits. They are
made available through Centre Donn\'ees de Strasbourg (CDS), and their format
and content are hereby described in brief.

%
%
\begin{table}
\caption[]{Description of the 2dF catalogue of $\omega$\,Centauri. The column
format is given in {\sc fortran} notation, where the columns are separated by
a single space.}
\begin{tabular}{lll}
\hline\hline
Column(s) & Format & Description \\
\hline
 1          & I5           & LEID (van Leeuwen et al.\ 2000) \\
 2\rlap{-4} & I2\ I2\ F5.2 & Right Ascension (J2000)
                             ($^{\rm h}\,^{\rm m}\,^{\rm s}$) \\
 5\rlap{-7} & I3\ I2\ F4.1 & Declination (J2000)
                             ($^\circ\,^\prime\,^{\prime\prime}$) \\
 8          & F9.5         & Galactic longitude ($^\circ$) \\
 9          & F8.5         & Galactic lattitude ($^\circ$) \\
10          & F6.3         & B (mag) \\
11          & F6.3         & B--V (mag) \\
12          & F6.3         & B$^\prime$ (mag) \\
13          & F6.3         & B$_1$--B$_2$ (mag) \\
14          & I4           & v$_{\rm LSR}$ (km s$^{-1}$) \\
15          & I5           & T$_{\rm eff}$ (K) \\
16          & F4.2         & log(g) (log cm s$^{-2}$)\\
17          & F5.2         & [Fe/H] \\
18          & I1           & Flag (see Section 4.1) \\
19          & I1           & If 1: two spectra were taken \\
20          & I1           & If 1: variable star of RR\,Lyrae-type \\
21          & F6.3         & S3839 \\
22          & F6.3         & S3933[blue] \\
23          & F6.3         & S3933[red] \\
24          & F6.3         & CH4300 \\
25          & F6.3         & Ba4554 \\
26          & F6.3         & TiO4620 \\
27          & F6.3         & TiO4760 \\
\hline
\end{tabular}
\end{table}

The catalogue has 27 columns (Table 3), and contains 1519 entries. The first
column lists the star's Leiden Identification number (LEID) as used in van
Leeuwen et al.\ (2000) and throughout this paper. Right Ascension and
Declination are given in the J2000 system in hours, minutes, seconds and
degrees, minutes, seconds notation. The Galactic coordinates are computed from
these in the subsequent two columns, and are given in degrees. The optical
photometry is taken from van Leeuwen et al.\ (2000) for $B$ and $B-V$, and
computed from the spectrum for $B^\prime$ and $B_1-B_2$ (see Section 4.1). The
radial velocity as measured from the spectrum is corrected for the movement of
the observatory with respect to the Sun and for the Sun's movement with
respect to the Local Standard of Rest (LSR). The effective temperature,
gravity and metallicity ([Fe/H] by proxy) are measured from the spectrum by
fitting {\sc atlas9} models (see Section 4.1). A flag indicates if a second
iteration was performed because possible emission was detected in the hydrogen
Balmer series or Ca H+K doublet (see Section 4.1). Another flag is set to 1 if
a second spectrum was taken, in which case all measurements are average
values. A third flag is set to 1 if the star is a known RR\,Lyrae-type
variable (van Leeuwen et al.\ 2000). The catalogue further lists the strengths
of the CN 3839 \AA\ band, interstellar components in the Ca\,{\sc ii} H+K
lines, CH 4300 \AA\ band, Ba\,{\sc ii} 4554 \AA\ line, and TiO 4620 and 4760
\AA\ bands (see Sections 5.2, 5.4.2, 5.7.1 and 5.7.3).

Spectra are made available for individual observations, where the file name
reflects the star's LEID number as well as the observation ID from the logbook
(Table 1). Each file has four columns and 1024 entries. The first column lists
the wavelengths in \AA, the second column lists the reduced 2dF spectrum
counts, the third column lists the normalised 2dF spectrum, and the fourth
column lists the best fitting {\sc atlas9}/{\sc synthe} spectrum (see Section
4.1).

\section{Results}

\subsection{Cluster membership and internal kinematics}

%
%
\begin{figure}
\centerline{\psfig{figure=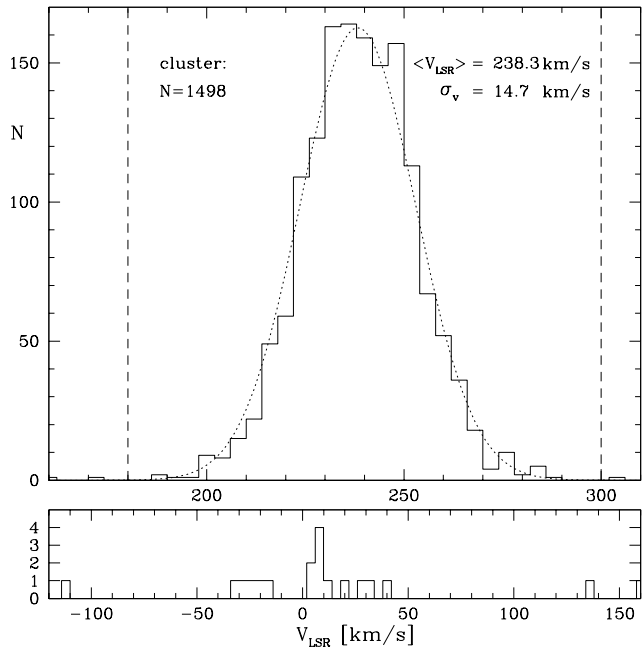,width=84mm}}
\caption[]{The velocity distribution clearly separates foreground disc stars
(bottom panel) from stars in $\omega$\,Cen (top panel), but this is less
obvious for a few stars just outside the bulk of cluster stars as delineated
by the vertical dashed lines --- these stars might be escaping from the
cluster, or they could be cluster binaries or halo stars (see text).}
\end{figure}

The large systemic radial velocity of $\omega$\,Cen compared to the foreground
Galactic disc facilitates the separation of field stars from cluster members
(Fig.\ 6). The radial velocities of the cluster members display a gaussian
distribution with a velocity dispersion of $\sigma_v=14.7$ km s$^{-1}$.
Correction for the velocity error of $\sigma_v=8$ km s$^{-1}$ (Section 4.2)
yields a true internal velocity dispersion of $\sigma_v=12.3$ km s$^{-1}$,
slightly smaller than the $\sigma_v=13.2$ km s$^{-1}$ determined by Pancino et
al.\ (2007) from spectra of a tenfold higher resolving power. The systemic
velocity $\langle v_{\rm LSR} \rangle = 238.3$ km s$^{-1}$ deviates by $+8$ km
s$^{-1}$ from the literature value (Meylan et al.\ 1995), which we attribute
to inaccurate absolute calibration of the 2dF wavelengths. Nonetheless, within
our set of measurements we define the realm of cluster members as $180<v_{\rm
LSR}<300$ km s$^{-1}$. The field stars are mostly found in the $-30<v_{\rm
LSR}<40$ km s$^{-1}$ region, but there are exceptions which may belong to the
halo, and it is possible that some field stars have radial velocities within
the range of $\omega$\,Cen.

Of the fifteen targets with $<90$\% proper motion membership probability, five
are confirmed to be field stars on the basis of their radial velocities. These
are: the RR\,Lyrae variable \#27076 ($v_{\rm LSR}=160$ km s$^{-1}$), the
$P=1.17$ d eclipsing binary \#37328, the late-M type star \#44420, the
short-period variable star \#60026 ($P=1.5$ hr), and \#78032. The other ten
are radial velocity members, and include the bright stars \#16018 and \#32029.

%
%
\begin{figure}
\centerline{\psfig{figure=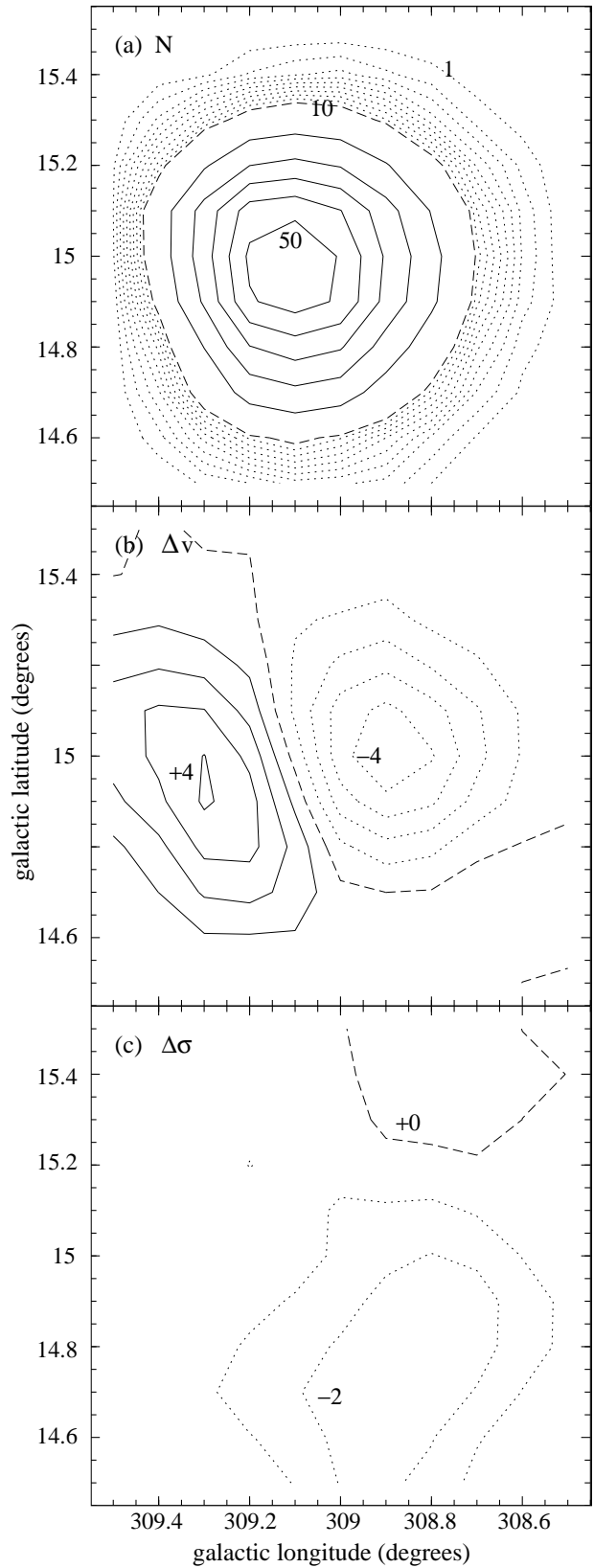,width=81mm}}
\caption[]{Sky distributions of: (a) the number of cluster members for which
2dF spectra are analysed, (b) the velocity with respect to the mean velocity
of $\langle v_{\rm LSR}\rangle =238.3$ km s$^{-1}$, and (c) the velocity
dispersion with respect to the mean velocity dispersion of $\sigma_{\rm
v}=14.7$ km s$^{-1}$. There is evidence for rotation in the inner part, but
the cluster periphery appears to be kinematically isothermal.}
\end{figure}

The internal kinematics of the cluster show rotation within $r<0.3^\circ$ at
an amplitude of $v_{\rm rot}\sim4$ km s$^{-1}$, with the axis of rotation
roughly aligned with the Galactic minor axis (Fig.\ 7b). This is somewhat
smaller than found previously by Meylan \& Mayor (1986), $v_{\rm rot}=8$ km
s$^{-1}$, or Reijns et al.\ (2006), $v_{\rm rot}=6$ km s$^{-1}$ (see also van
de Ven et al.\ 2006), and in part due to the angular resolution of $0.1^\circ$
employed in constructing the maps. The velocity dispersion shows no evidence
of spatial variations in excess of 2 km s$^{-1}$ out to $r\sim0.5^\circ$
(Fig.\ 7c), but this is due to the limited accuracy of the velocity
measurements as well as the $\sim0.1^\circ$ angular resolution in the core.
Hence we are unable to detect the radial gradients in velocity dispersion
reported in the literature (Meylan \& Mayor 1986; Norris et al.\ 1997; Reijns
et al.\ 2006).

\subsection{Interstellar absorption by ionised gas}

%
%
\begin{figure}
\centerline{\psfig{figure=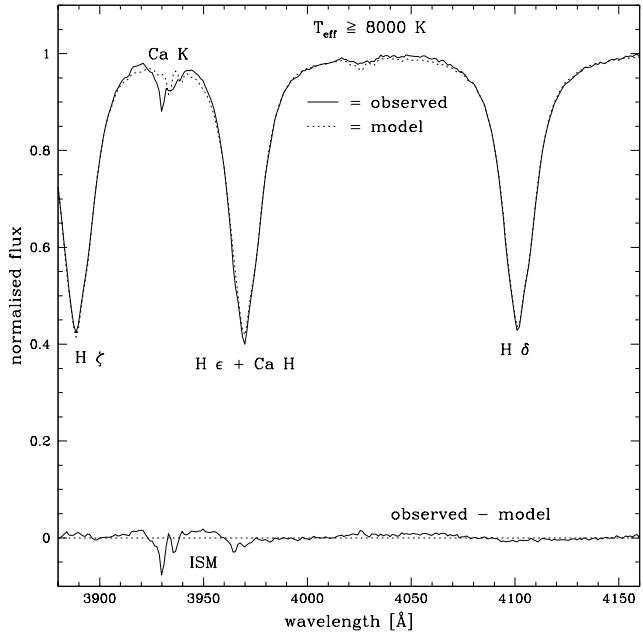,width=84mm}}
\caption[]{Average of 263 normalised spectra of 231 hot stars, with $T_{\rm
eff}\geq8000$ K. The average best model (dotted line) fits the Balmer lines
perfectly, clearly revealing the interstellar (ISM) component in the Ca\,{\sc
ii} H and K lines; because the stellar spectra have been Doppler-corrected,
the ISM components appear shifted with respect to the rest wavelengths of 3968
and 3933 \AA.}
\end{figure}

In hot stars, with $T_{\rm eff}\geq8000$ K, the photospheric Ca\,{\sc ii} H+K
lines at 3968 and 3933 \AA\ are sufficiently weak for the ISM components to be
reliably measured (Fig.\ 8). The ISM component in the K line can be discerned
in metal-poor stars as cool as $T_{\rm eff}\sim6000$ K, but it is difficult to
obtain accurate measurements. The same is true for the H line even in hot
stars because there it blends with the very strong H$\epsilon$ line. Of the
classified spectra, 263 are of 231 unique hot stars. Their average not only
shows both ISM components very clearly, but each seems to be split in a strong
blue-shifted component and a weaker {\it red}-shifted component (Fig.\ 8). The
former is entirely consistent with the expected foreground ionised gas in the
Galactic disc, but the presence of gas moving at a speed of $\gsim100$ km
s$^{-1}$ towards the cluster is unexpected.

%
%
\begin{figure}
\centerline{\psfig{figure=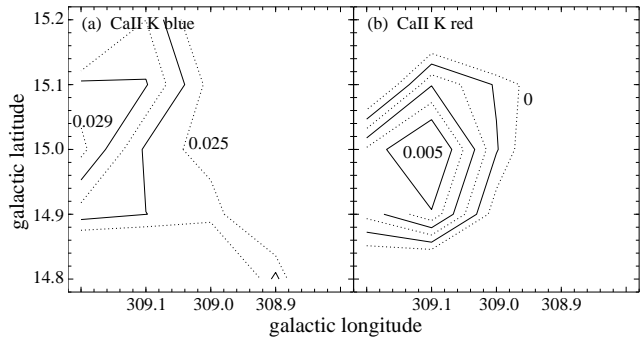,width=84mm}}
\caption[]{Spatial distribution of ISM absorption in the Ca\,{\sc ii}\,K line.
The blue-shifted component (a) shows a weak gradient across the face of the
cluster, whilst the red-shifted component (b) is concentrated near the cluster
centre.}
\end{figure}

We measured the strength of the blue- and red-shifted Ca\,{\sc ii}\,K ISM
components from the average flux levels, $F_{\lambda1-\lambda2}$ (between
wavelengths $\lambda1$ and $\lambda2$ in \AA), in the normalised spectrum
after subtracting the best model fit, in the following way:
\begin{equation}
S3933[{\rm blue}] = F_{3921-3932} - \frac{ F_{3916-3921} + F_{3932-3934} }{2},
\end{equation}
and
\begin{equation}
S3933[{\rm red}] = F_{3934-3939} - \frac{ F_{3932-3934} + F_{3939-3950} }{2}.
\end{equation}
Their variations across the sky are different (Fig.\ 9), with the blue-shifted
component displaying a weak gradient across the face of the cluster, whilst
the red-shifted component is clearly concentrated near the cluster centre.
This confirms that the blue-shifted component is due to absorption in ionised
gas within the Galactic disc, but it hints at the red-shifted component to be
associated with gas within the gravitational influence of the cluster.

\subsection{Metallicity and the morphology of the Hertzsprung-Russell Diagram}

%
%
\begin{figure}
\centerline{\psfig{figure=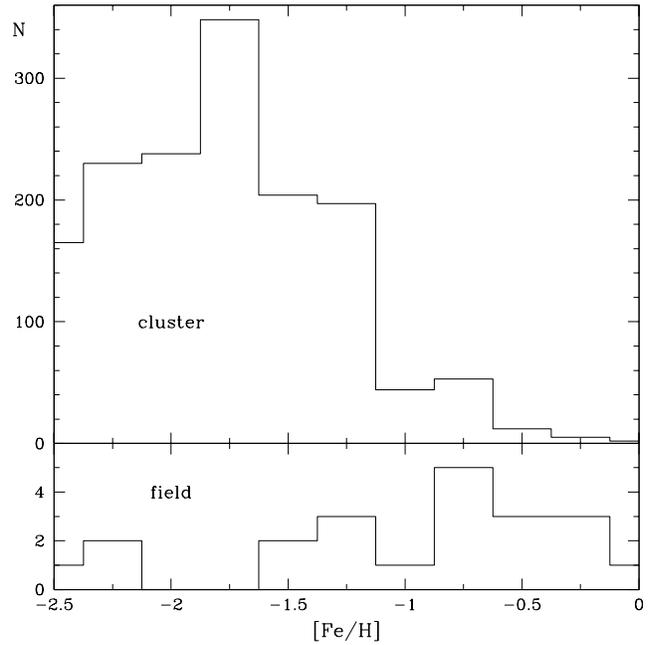,width=84mm}}
\caption[]{The metallicity distribution of $\omega$\,Cen (top panel) peaks at
[Fe/H]$\simeq-1.8$, but the spread is large, $\Delta$[Fe/H]$\sim1$, and a
distinct sub-population around [Fe/H]$\simeq-0.8$ is detected. The foreground
stars, on the other hand, are mostly found to have [Fe/H]$>-1.5$ (bottom
panel).}
\end{figure}

The metallicity distribution of cluster members clearly peaks around
[Fe/H]$\simeq-1.8$, but the bulk of the stars are spread over a range between
[Fe/H]$\sim-2.3$ to [Fe/H] $\sim-1.3$, with a small distinct peak at
[Fe/H]$\simeq-0.8$ (Fig.\ 10). This metal-rich sub-population has spatial and
radial velocity distributions which are indistinguishable from those of the
bulk of $\omega$\,Cen members. The uncertainties in the [Fe/H] measurements do
not allow to identify more sub-populations. The field stars generally have
higher metallicities than the cluster members, as expected for the foreground
Galactic disc populations, but some metal-poor stars are found in the field
and these may belong to the Galactic halo.

%
%
\begin{figure}
\centerline{\psfig{figure=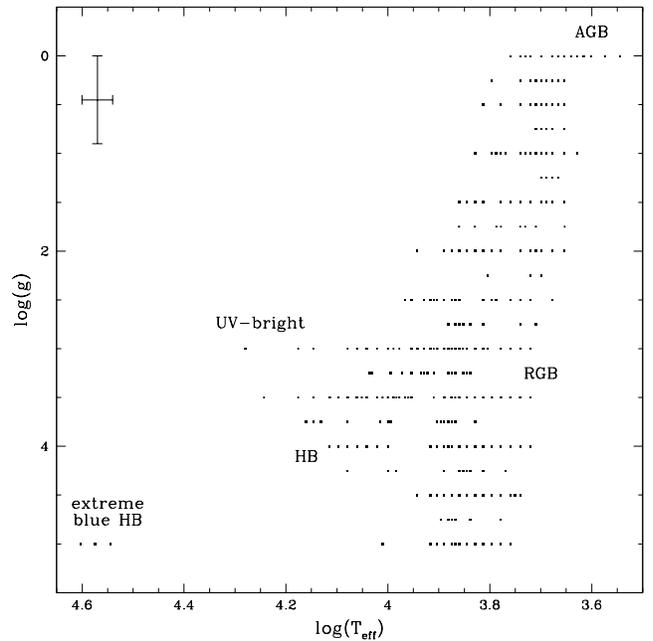,width=84mm}}
\caption[]{Physical Hertzsprung-Russell Diagram, where gravity (a luminosity
indicator) is plotted versus effective temperature. The typical measurement
uncertainty is indicated in the top left.}
\end{figure}

The measurement of gravity and temperature enables us to construct a {\it
physical} Hertzsprung-Russell Diagram (Fig.\ 11), as opposed to a
colour-magnitude diagram. Although the typical measurement error on individual
stars is quite substantial (Table 2), the stars nevertheless clearly outline
the RGB and HB both in terms of their gravities and temperatures. The gravity
is mainly an indicator of stellar radius (and thus luminosity), as the
post-main-sequence stars in $\omega$\,Cen have similar masses: whilst the
luminosities span two orders of magnitude, the masses only vary from $\sim0.8$
M$_\odot$ for RGB stars that have yet to undergo significant mass loss down to
$\sim0.5$ M$_\odot$ for post-AGB and UV-bright stars that will cool to become
white dwarfs. As expected, the coolest and largest stars are found near the
tips of the RGB and AGB, whilst the extreme blue HB stars are hot and compact.

%
%
\begin{figure*}
\centerline{\psfig{figure=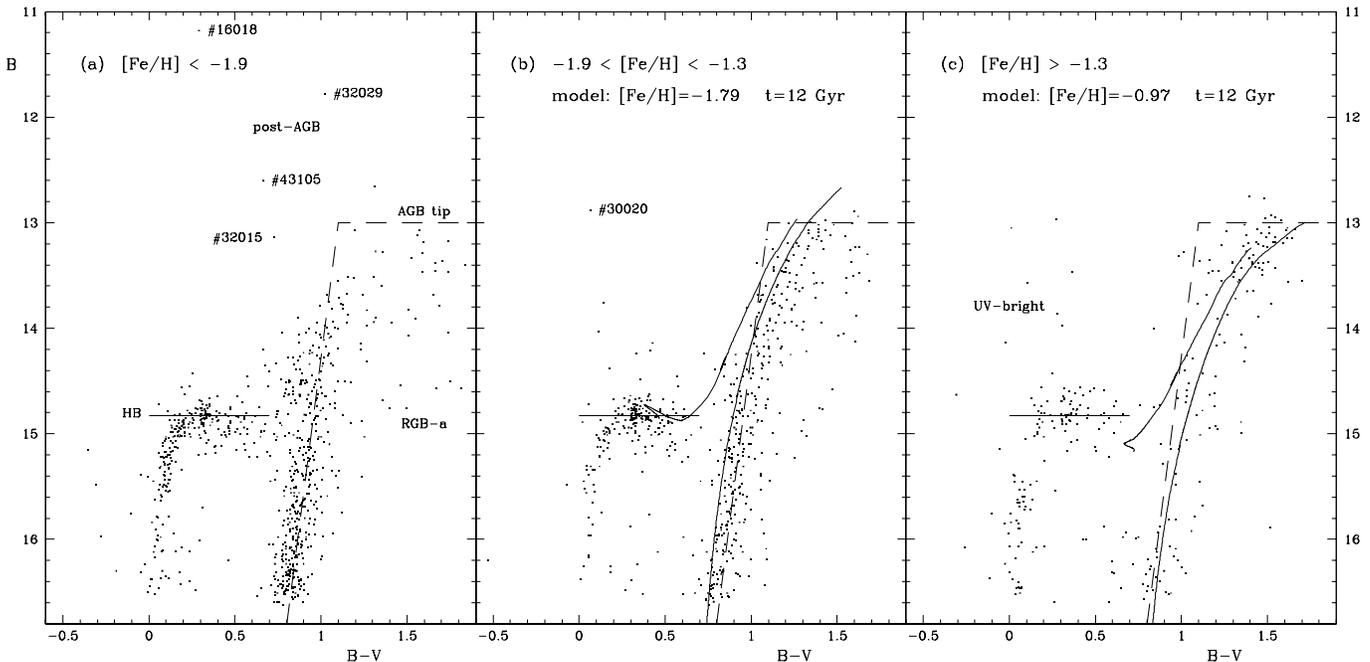,width=180mm}}
\caption[]{Colour-magnitude diagrams of $\omega$\,Cen members, for different
metallicities. The dashed lines are for reference, delineating the lower RGB
and the location of the tip of the AGB; similarly, the solid line indicates
the B-magnitude of the RR\,Lyrae stars (the clump of stars especially clearly
visible in the intermediate-metallicity stars in panel b). Theoretical
isochrones from Pietrinferni et al.\ (2004) are plotted for [Fe/H]$=-1.79$ and
$-0.97$, respectively, for an age of 12 Gyr, a distance of 5.2 kpc, and
reddened by $E(B-V)=0.11$ mag (see text for discussion). A few remarkable
post-AGB star candidates are labelled.}
\end{figure*}

In order to investigate the dependence of the late stages of stellar evolution
on metallicity, the (B, B--V) colour-magnitude diagram is constructed for
cluster members in three different metallicity bins: metal-poor, [Fe/H]$<-1.9$
(Fig.\ 12a), intermediate, $-1.9<$[Fe/H]$<-1.3$ (Fig.\ 12b), and metal-rich,
[Fe/H]$>-1.3$ (Fig.\ 12c). The differences are quite subtle. Individual types
of stars are discussed in more detail in subsequent sections.

It seems that the most luminous post-AGB stars are metal-poor, whilst the
UV-bright stars are generally more metal-rich (Fig.\ 12). This may be related
to the timescale of evolution of the post-AGB stars, in particular if
UV-bright stars are more evolved post-AGB stars. If, on the other hand,
UV-bright stars are AGB manqu\'e stars, i.e.\ stars which evolve from the HB
directly towards the white dwarf cooling track bypassing the AGB stage
(Greggio \& Renzini 1990), then this would imply that the brightest metal-rich
stars are on the RGB rather than on the AGB, or in other words the AGB would
be dominated by less metal-rich stars.

The clump of stars around $B\sim14.83$ and $B-V\sim0.3$ to 0.4 mag (Fig.\ 12)
are mostly RR\,Lyrae variables; the absence of such clump at [Fe/H]$>-1.3$ is
commensurate with the notion of RR\,Lyrae variability being a feature of
metal-poor populations. Metal-rich stars do straggle above the HB (around
$B\sim14.6$ mag) more than their metal-poor siblings. Also, the (extreme) blue
HB does not seem to be an exclusive feature of the most metal-poor stars,
although the poor statistics at [Fe/H]$>-1$ hampers deriving a more definitive
conclusion.

Theoretical isochrones from Pietrinferni et al.\ (2004) are plotted in Fig.\
12 for [Fe/H]$=-1.79$ and $-0.97$, respectively. Their standard model without
convective overshooting and with solar-scaled abundances is used, for an age
of 12 Gyr (cf.\ Villanova et al.\ 2007; Stanford et al.\ 2006a), a distance of
5.2 kpc (cf.\ Del Principe et al.\ 2006; van de Ven et al.\ 2006; Caputo et
al.\ 2002), and reddened by $E(B-V)=0.11$ mag (Lub 2002). There is rough
correspondance between the isochrones and the location of the RGB although the
metal-poor model is somewhat too bright. The metal-poor model produces stars
near the location of the RR\,Lyrae instability strip, but the bluer part of
the HB is not reproduced. Neither is there evidence in the observed (B, B--V)
diagram for the existence of a red clump, predicted for [Fe/H]$\gsim-1$. This
demonstrates the difficulty of theoretical models to reproduce all features
seen in globular cluster photometric diagrams.

Surprisingly, the anomalous RGB branch (RGB-a) turns out to be metal-poor,
much in contrast with the usual interpretations of their red optical colours
and deep calcium absorption lines. This is investigated in more detail in the
next section.

\subsection{The first ascent Red Giant Branch (RGB)}

Because of the sheer numbers of bright stars, the RGB is a popular place to
disect the stellar populations in globular clusters. The spread in optical
colours and magnitudes near the tip of the RGB and the much fainter anomalous
branch (RGB-a) are generally considered to be the result of multiple stellar
populations, which are expected to show differences in overall metallicity
and/or certain elemental abundances.

Some of the fainter stars ($B\sim16$ mag) have B--V colours in between those
of RGB stars ($B-V\,\gsim\,0.7$ mag) and those of blue HB stars
($B-V\,\lsim\,0.2$ mag). We attribute this to photometric errors, as their
spectra and temperatures leave no doubt about them being normal RGB stars.

\subsubsection{The nature of the anomalous RGB branch}

The RGB of metal-rich stars is offset to redder B--V colours by about 0.1 to
0.2 magnitudes (Fig.\ 12), as expected if their temperatures are slightly
reduced because of the enhanced opacity in their mantles and the consequently
larger stellar radii. However, the RGB-a does {\it not} seem to be metal-rich.
Selecting the 13 stars with $14.4<B<15.1$ and $B-V>1.3$ mag, these have an
average metallicity much in line with the bulk of the $\omega$\,Cen
population: $\langle$[Fe/H]$\rangle_{\rm RGB-a}=-1.76\pm0.43$. However, their
temperatures are much lower: $\langle\log(T_{\rm eff})\rangle_{\rm
RGB-a}=3.59\pm0.04$ (3867 K). In all but one case, the gravity was determined
to be $\log(g)=0$, which places these stars near the tip of the AGB in the
physical HRD (Fig.\ 11). If they are in fact AGB stars, the $\sim$1.6 mag
difference in B-band magnitude would na\"{\i}vely suggest that they lie
roughly a factor two further away than $\omega$\,Cen, placing them in the
outskirts of the Galactic bulge. On the other hand, their radial velocities,
proper motions and locations on the sky all point at locations within the
cluster (van Leeuwen et al.\ 2000; Pancino et al.\ 2002). It is much more
likely therefore that they do have the bolometric magnitudes of brighter red
giants in $\omega$\,Cen, but that their lower temperatures yield much greater
bolometric corrections to their B-band magnitudes. Indeed, RGB-a stars appear
much brighter than other cluster stars in the near-infrared than at optical
wavelengths (Sollima et al.\ 2004). Bolometric corrections to optical
magnitudes of stars near the tip of the AGB are also relatively large as they
have a very low surface gravity (Fig.\ 5): they are cool and luminous and
likely less massive than RGB stars due to mass loss. Therefore, it is possible
that some bright AGB stars reach the region of the optical CMD populated by
RGB-a stars. We must stress the difficulty of measuring very cool stars
($T<4000$ K), in particular in distinguishing between a low temperature or a
high metallicity.

%
%
\begin{figure}
\centerline{\psfig{figure=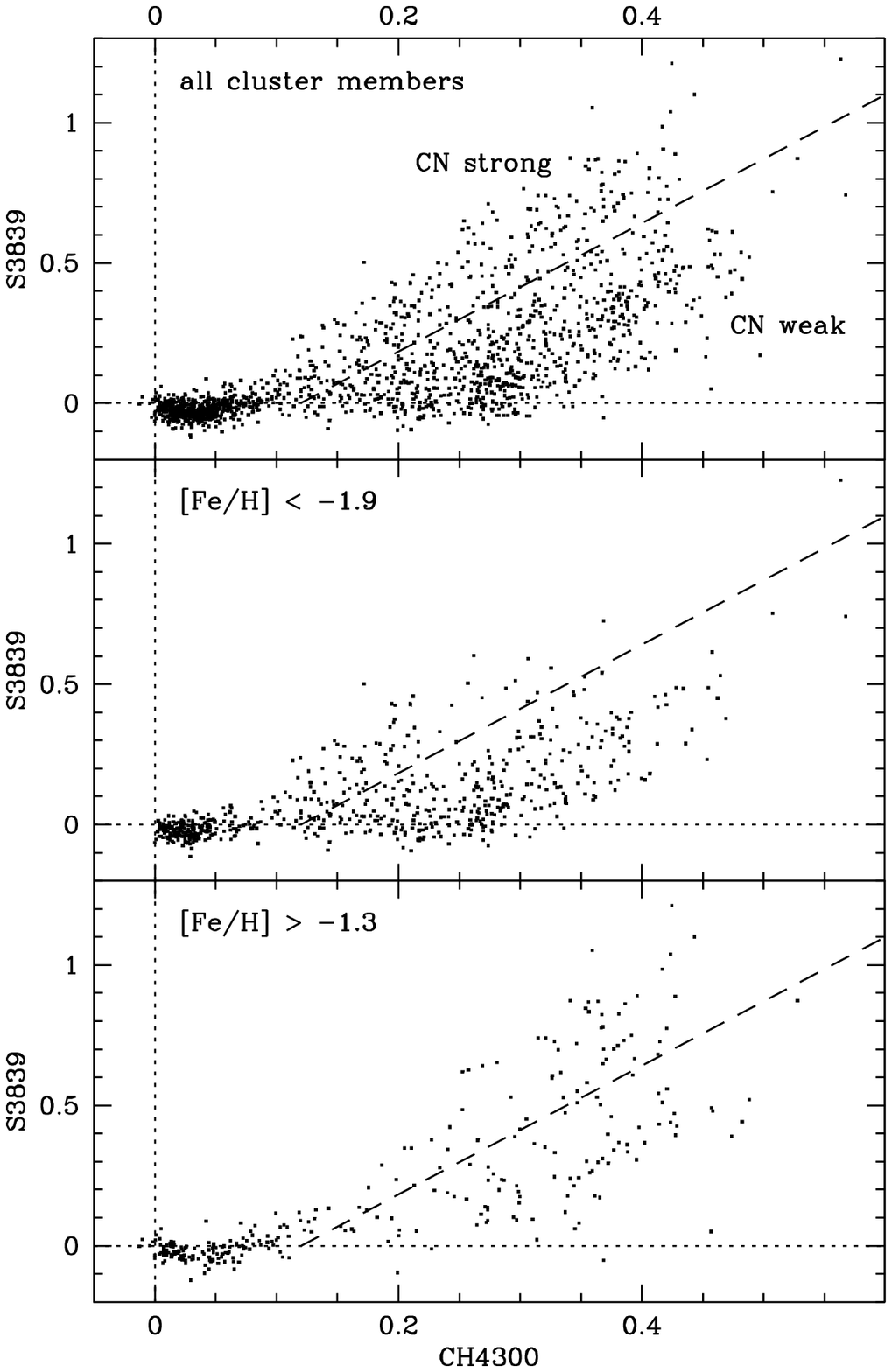,width=84mm}}
\caption[]{The strength of the 3839 \AA\ band of CN versus that of the 4300
\AA\ (G-)band of CH. For cool stars with a strong G-band, there is a dichotomy
in the strength of the CN band.}
\end{figure}

\subsubsection{Nitrogen enrichment}

We measured the strength of the strong CN band with a bandhead at 3839 \AA,
and the strength of the CH band at 4300 \AA\ (the G-band) from the average
flux levels, $F_{\lambda1-\lambda2}$ (between wavelengths $\lambda1$ and
$\lambda2$ in \AA), in the normalised spectrum after subtracting the best
model fit, following Harbeck et al.\ (2003):
\begin{equation}
S3839 = -2.5 \log\left( \frac{F_{3861-3884}}{F_{3894-3910}} \right),
\end{equation}
and
\begin{equation}
CH4300 = -2.5 \log\left( \frac{2 F_{4285-4315}}{F_{4240-4280}+F_{4390-4460}}
\right).
\end{equation}
The G-band is visible at higher temperatures than the CN band, which only
becomes visible once $CH4300\,\gsim\,0.1$ but this is not guaranteed to be the
case until $CH4300\,\gsim\,0.35$ (Fig.\ 13). There is a clear dichotomy
between CN-weak and CN-strong stars, at a given strength of the G-band. This
must be due to differences in the nitrogen abundance, as often the
temperature, gravity and overall metallicity are the same between stars with
and without a strong CN band. However, a weak correlation with overall
metallicity is present, with the CN-strong stars being more often metal-rich
than the CN-weak stars (Fig.\ 13). In particular, the strongest CN stars with
$S3839>0.6$ are almost exclusively metal-rich, with [Fe/H]$>-1.3$. This is in
agreement with the results from Str\"omgren photometry by Hilker \& Richtler
(2000) and the recent work by Kayser et al.\ (2006) on less evolved stars.

%
%
\begin{figure}
\centerline{\psfig{figure=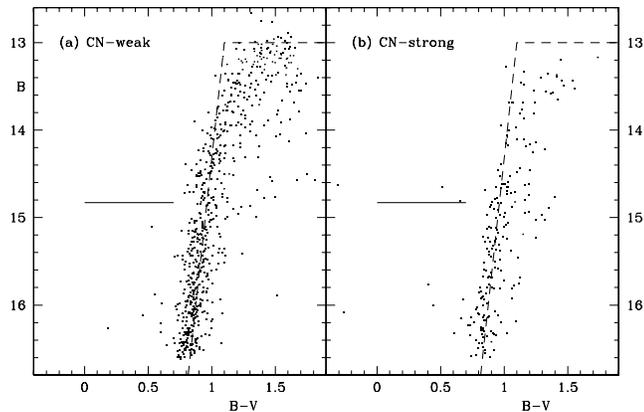,width=84mm}}
\caption[]{Colour-magnitude diagram of stars with a clear G-band,
$CH4300>0.15$, split into (a) CN-weak stars and (b) CN-strong stars. These are
all red giants, with the odd ``blue'' star being there due to erroneous
photometry.}
\end{figure}

The CN strength shows very little correlation with the position in the
colour-magnitude diagram, although it is clear that the nature of RGB-a is
{\it not} related to nitrogen enrichment (Fig.\ 14). Also, the CN-strong stars
seem to avoid the tip of the AGB, which is somewhat surprising given that the
tip of the AGB has a relatively larger number of metal-rich stars (Fig.\ 13).
As for the RGB-a stars, this can be explained if the CN-strong stars are
somewhat cooler, yielding greater bolometric corrections to the B-band.

\subsubsection{Chromospheric activity}

%
%
\begin{figure}
\centerline{\psfig{figure=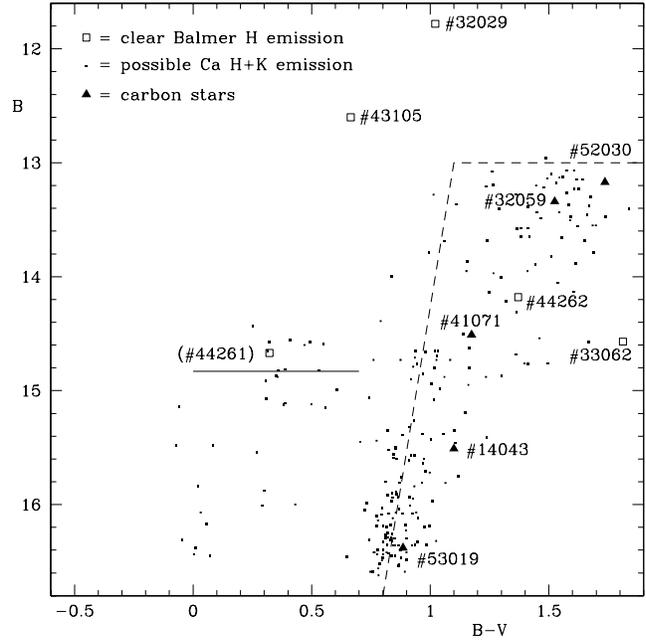,width=84mm}}
\caption[]{Colour-magnitude diagram of stars with hydrogen line emission
(squares), stars with possible Ca\,H+K emission (dots), and carbon stars
(triangles). The H emitters and carbon stars are labelled individually as they
are so few --- \#44261 in fact suffered from contaminating light from the very
strong H emitter \#44262 (V42) at a distance of $5^{\prime\prime}$.}
\end{figure}

In chromospherically active stars the Ca\,{\sc ii}\,H+K lines show line-core
inversion due to an emission component (Dupree \& Smith 1995). At the moderate
resolution of our 2dF spectra, these inversions will not be resolved but
instead lead to filling-in of the absorption core. The flag in the catalogue
(Table 3, Section 4.1) was used as a criterion to isolate possibly
chromospherically active stars (flag=2). Their locations in the
colour-magnitude diagram indicate that they are predominantly found amongst
red giants (Fig.\ 15). Indeed, the warmer stars near the bottom of the RGB are
expected to have more prominent chromospheres than the AGB stars that rotate
slowly or than the HB stars that do not have deep convective mantles. In
particular, none of the post-AGB or UV-bright stars are flagged. However, the
lower signal at the bottom of the RGB would have led to more spurious flag=2
detections in any case, due to the larger inaccuracies in the absorption line
fitting and/or sky background subtraction. In the same way we discard the blue
HB stars with flag=2 due to the lower S/N ratio in those faint stars. It is
then remarkable that quite many stars near the tip of the RGB and/or AGB are
flagged. Their chromospheres may be heated by mechanical energy deposition.
Alternatively, it may be too difficult to accurately model the strong Ca\,{\sc
ii}\,H+K absorption lines in these cool, low-gravity stars.

\subsection{The RR\,Lyrae instability strip}

%
%
\begin{figure}
\centerline{\psfig{figure=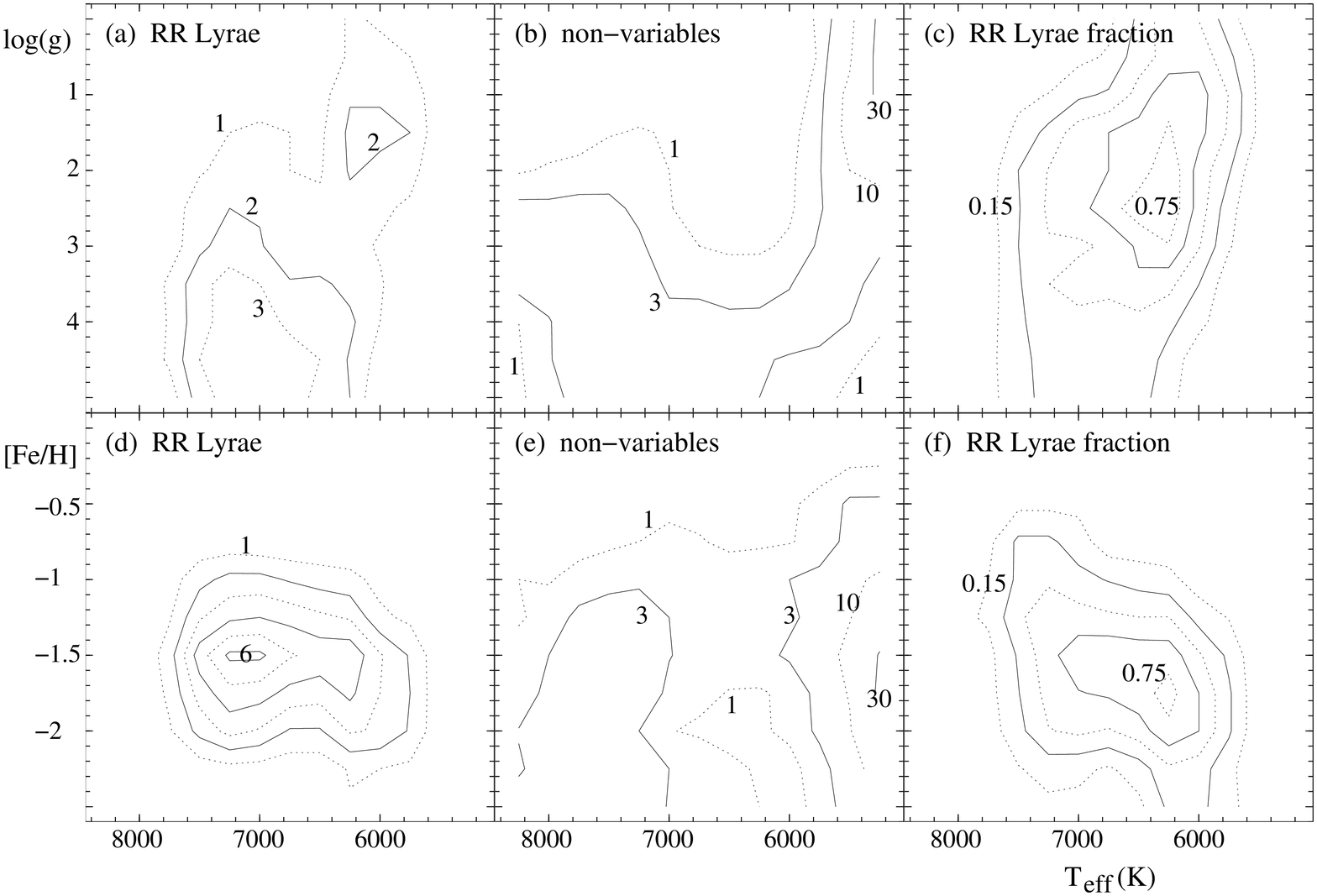,width=84mm}}
\caption[]{Temperatures, gravities and metallicities of RR\,Lyrae variables
(both in absolute numbers and as a fraction of the total number of stars) and
non-variables in $\omega$\,Centauri.}
\end{figure}

The RR\,Lyrae variables in $\omega$\,Cen have a typical temperature of $T_{\rm
eff}=7000$ K, but a small number of cooler RR\,Lyrae variables exist with
$T_{\rm eff}=6000$ K (Fig.\ 16a). They also differ in gravity, the cooler
atmospheres being significantly lighter. They do not, however, differ in
metallicity, peaking around the cluster's bulk metallicity (Fig.\ 16d). The
cooler variables may be on the way up the AGB. There are fewer non-variable
stars between 6000 and 7000 K, but there is significant overlap with
especially the warm RR\,Lyrae (Figs.\ 16b and 16e), whilst a large fraction of
the stars around $T_{\rm eff}=6000$ K and $\log(g)=1$ to 2 seem to vary (Fig.\
16c). The cooler stars are more likely to vary if they are of low metallicity,
whilst the opposite is the case for the warmer variables up to $T_{\rm
eff}\sim7500$ K (Fig.\ 16f).

We have spectroscopic variability information for 45 RR\,Lyrae variables.
Although this only comprises one repeat per such star, the time interval is
comparable to or larger than their periods and it therefore allows us to make
statistical statements about the amplitudes of variability. The differences in
stellar parameters deduced from the two epochs are identical to those deduced
for other, non-variable stars, except for the variation in gravity which is
with $\Delta\log(g)=1.38$ much larger than that of any other type of star (see
Section 4.2). This may be the result of the movement and compression/dilution
of the atmosphere in the radial pulsation cycle. Comparison with lightcurves
(Weldrake et al.\ 2007) may reveal phased behaviour of the gravity.

The 49 non-variable stars in the RR\,Lyrae box (Section 3) for which repeat
observations are available still show a relatively large dispersion of
$\Delta\log(g)=1.05$; these stars may be pulsating too, albeit at a lower and
hitherto unnoticed level. It is interesting to note that surface-level
pulsation modes at milli-mag amplitudes have been predicted to occur in stars
at the warm edge of the RR\,Lyrae/Cepheid instability strip, by Buchler \&
Koll\'ath (2001).

%
%
\begin{figure}
\centerline{\psfig{figure=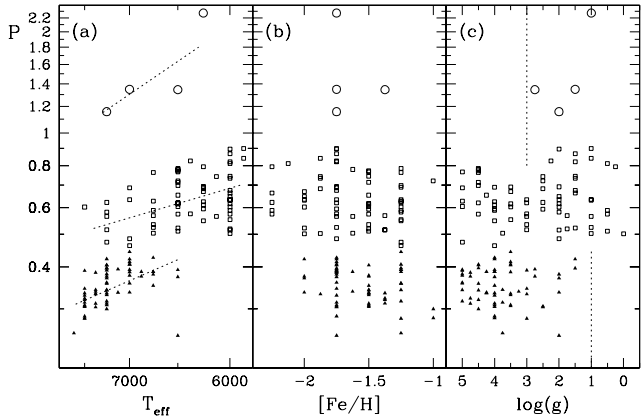,width=84mm}}
\caption[]{Distributions of RR\,Lyrae variables in $\omega$\,Cen over their
pulsation periods, as a function of effective temperature, metallicity, and
surface gravity. Different symbols are used for stars pulsating in the first
overtone ($P<0.45$ d; filled triangles) or at longer periods ($P>1$ d; open
circles). The dotted lines are meant to guide the eye (see text).}
\end{figure}

The RR\,Lyrae variables can be divided into RRab-type which pulsate in the
fundamental mode at periods between $P\sim0.45$ and 1 day, overtone RRc-type
pulsators with $P<0.45$ d, and longer-period RR\,Lyrae variables with periods
in excess of a day (Fig.\ 17). The period increases at lower temperature in
all three classes of variables, but at a different rate (dotted lines in Fig.\
17a). RRab-types are cooler than RRc-types by $\sim500$ K on average. The RRab
and RRc-types do not separate in metallicity, although the average metallicity
of the RRc-types is slightly higher than that of RRab-types (Fig.\ 17b).
RRab-types and RRc-types span a range in gravities, but the RRab-types reach
lower gravities than the RRc-types, and the longer-period RR\,Lyrae variables
also tend to have low gravities (Fig.\ 17c). This can be understood in terms
of the larger radii of longer-period RR\,Lyrae variables.

%
%
\begin{figure}
\centerline{\psfig{figure=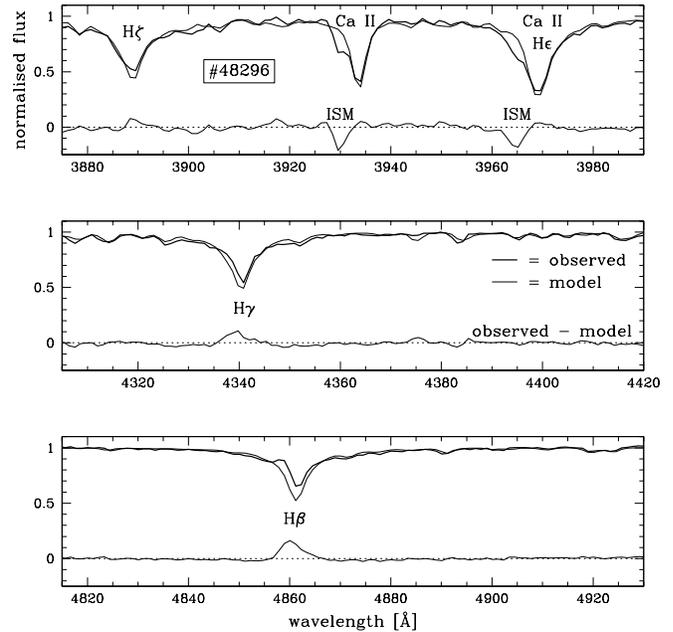,width=84mm}}
\caption[]{The spectrum of RRa-type variable star \#48296. Besides very strong
interstellar absorption lines of Ca\,{\sc ii} it also shows blueshifted Balmer
line emission.}
\end{figure}

The RRa-type variable star \#48296 shows blue-shifted Balmer line emission
(Fig.\ 18). This star belongs to the cool class of RR\,Lyrae, and has an
unremarkable pulsation period of $P=0.663$ days. Its radial velocity is with
$v_{\rm LSR}=255$ km s$^{-1}$ fairly high but well within the velocity
envelope of the cluster (Fig.\ 6). It is located at Galactic coordinates
$b=309.1372^\circ$ and $l=14.9253^\circ$, towards the region of generally high
interstellar absorption (Fig.\ 9). There is a bright 24-$\mu$m source at a
projected distance of just $10^{\prime\prime}$ (Boyer et al.\ 2007). The
nature of the Balmer line emission could be due to shocks in the pulsating
atmosphere and possibly indicate mass loss.

\subsection{Extreme Horizontal Branch stars (EHB)}

%
%
\begin{figure}
\centerline{\psfig{figure=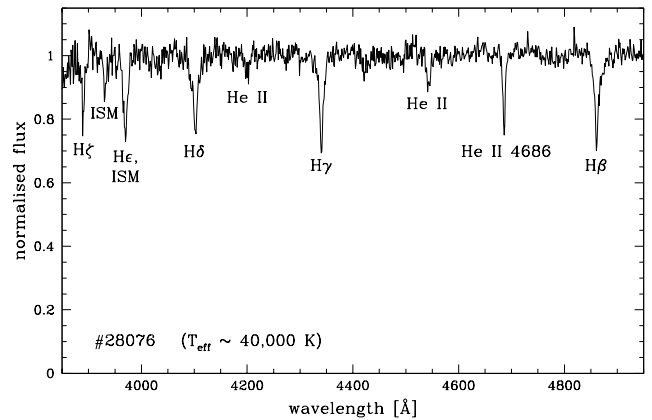,width=84mm}}
\caption[]{Spectrum of the hottest star detected in our survey, \#28076
($T_{\rm eff}\sim40,000$ K). The prominence of the He\,{\sc ii} 4686 line is
clear evidence for its extremely high temperature.}
\end{figure}

The HB extends to very high temperatures. Although the bulk of HB stars have
$T_{\rm eff}<15,000$ K, we detected six stars with $T_{\rm eff}\geq15,000$ K,
of which two stars have $T_{\rm eff}>20,000$ K. All of them appear to have
[Fe/H]$\geq-1.25$, although there may be systematic uncertainties because the
metallicity relies on reproducing the exact shapes of the hydrogen Balmer
lines due to the lack of metallic absorption lines at those high temperatures.
The metallicity measurements may also be affected by radial gradients within
the atmosphere due to radial levitation, diffusive sedimentation, et cetera
(Behr et al.\ 1999). By far the hottest stars are \#27010 ($T_{\rm
eff}=37,500$ K) and \#28076 ($T_{\rm eff}=40,000$ K). Both stars are assigned
the highest value for the gravity, $\log(g)=5$. They are quite faint as the
bolometric corrections to their B-band magnitudes are substantial, and in the
case of \#27010 this may have led to a slight overestimation of the
temperature. The spectrum of \#28076 leaves no doubt about its high
temperature, displaying strong He\,{\sc ii} lines (Fig.\ 19) --- these are not
included in the models because at such high temperature non-Local
Thermodynamic Equilibrium effects become important. It is interesting to note,
though, that helium enrichment can explain the high temperatures of EHB stars
(Lee et al.\ 2005).

\subsection{The Asymptotic Giant Branch (AGB)}

How can we isolate the AGB stars from the much more numerous RGB stars? This
is an important question if we are to understand which stars survive the
core-helium burning phase and undergo thermal pulses and $3^{\rm rd}$
dredge-up. To identify and investigate the AGB stars in $\omega$\,Cen, we look
for the coolest stars (Section 5.7.1), long-period variables (Section 5.7.2)
and the chemical signatures of dredge-up (Section 5.7.3, where the special
case of carbon stars is discussed separately in Section 5.8).

\subsubsection{M-type stars}

The M type spectral class is defined on the basis of the appearance of TiO
bands in the optical spectrum (e.g., Jaschek \& Jaschek 1990). The rather
abrupt temperature threshold below which TiO forms, the ease with which it
forms, the relatively high abundance of oxygen and titanium in most stars
(notable exceptions include carbon stars, see Section 5.8), the low excitation
of the TiO band transitions and their occurrence throughout the optical
spectrum makes M-type stars amongst the easiest to identify. In spite of this,
there appear to be only a handful of M-type stars in our sample. The question
is whether these are AGB stars, and in what way they are distinct from the
other cool giant stars such as the CN-rich stars. With regard to overall
metallicity the M-type stars are not particularly metal-rich; only \#44262
(V42) may have a somewhat higher metallicity at an estimated [Fe/H]$=-1.25$.

%
%
\begin{figure}
\centerline{\psfig{figure=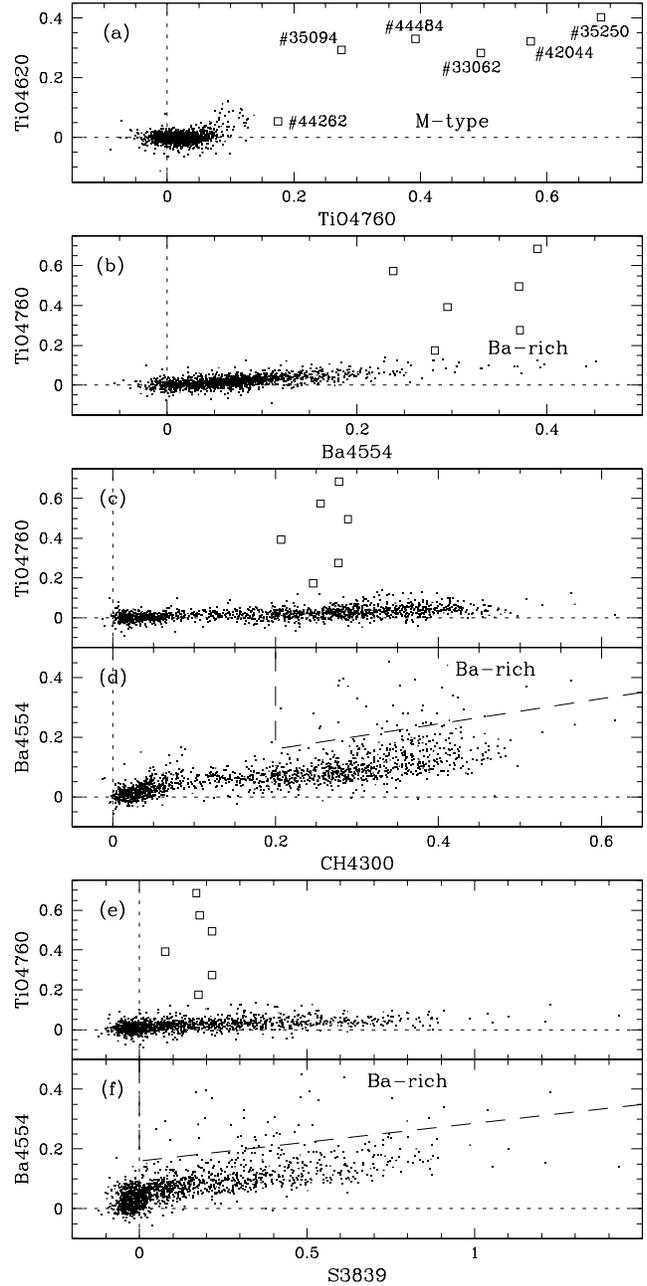,width=84mm}}
\caption[]{Comparisons between the strength of the TiO and barium absorption,
and that of the CH and CN absorption. The M-type stars are indicated with
squares and labelled.}
\end{figure}

To measure the strength of the TiO bands we define two indices, the first of
which measures the prominence of the $\lambda$4760 bandhead:
\begin{equation}
TiO4760 = -2.5 \log\left( \frac{F_{4764-4766}}{F_{4754-4756}} \right),
\end{equation}
and the second of which measures the strength of absorption at either side of
the pseudo-continuum around 4620 \AA:
\begin{equation}
TiO4620 = -2.5 \log\left( \frac{F_{4590-4592}+F_{4640-4642}}{2 F_{4616-4620}}
\right),
\end{equation}
where the average flux levels, $F_{\lambda1-\lambda2}$ (between wavelengths
$\lambda1$ and $\lambda2$ in \AA), are measured in the normalised spectrum.
There are a few dozen stars in our sample with detectable TiO absorption, at
$TiO4760>0.07$ and $TiO4620>0.02$ (Fig.\ 20a). Six stars clearly stand out
from the rest, with TiO absorption at a 22 per cent level or more except
\#44262 (V42) with a TiO 4760 \AA\ bandhead of 8.5 per cent of the continuum;
these are identified with different symbols and labelled. While the 4620 \AA\
feature seems to saturate around 30 per cent ($TiO4620\sim0.3$ to 0.4), the
TiO 4760 \AA\ band can reach a depth of more than half the continuum --- the
most extreme example in our sample is \#44420 with indices $TiO4760=0.92$ and
$TiO4620=0.40$ (not included in Fig.\ 20 as it is a foreground star).

\subsubsection{Pulsation}

%
%
\begin{figure}
\centerline{\psfig{figure=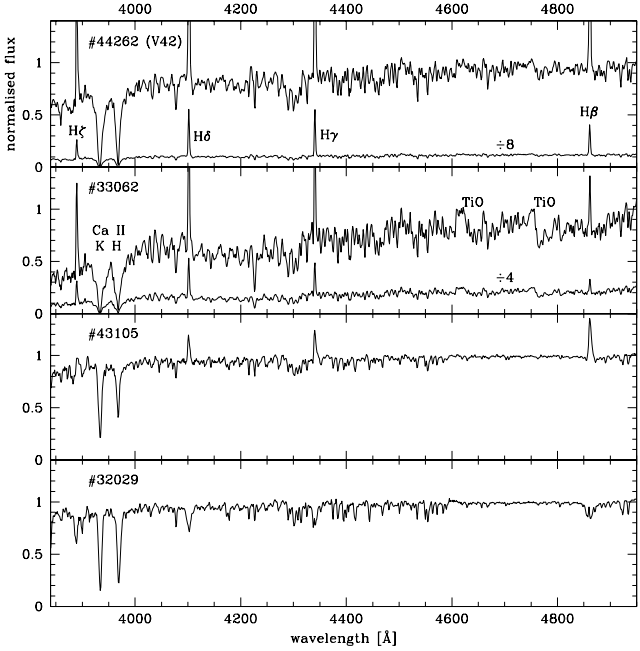,width=84mm}}
\caption[]{Spectra of the hydrogen Balmer line emission stars detected in our
survey. These include the two M-type AGB stars \#44262 (V42) and \#33062, and
the two bright post-AGB stars \#43105 and \#32029.}
\end{figure}

The coolest and most luminous AGB stars are the most likely stars to undergo
strong radial pulsations on timescales of months to years, resulting in large
variations in brightness. Indeed, with a period of $P=149$ days (Sawyer Hogg
1973; Clement 1997; van Leeuwen et al.\ 2000) the longest-period variable star
in $\omega$\,Cen, V42 (\#44262) is one of the few M-type stars in our sample
(albeit not the coolest). It also exhibits extremely strong Balmer line
emission indicative of the strong shocks that develop in the radially
pulsating atmosphere (Fig.\ 21), shown in detail in a high-resolution spectrum
by McDonald \& van Loon (2007). The coolest M-type star in our sample, \#35250
is also a known long-period variable ($P=51$ days), but none of the other
M-type stars are. Nonetheless, the cool M-type star \#33062 is likely to
pulsate as it is the only star in our sample besides V42 with strong Balmer
line emission (Fig.\ 21).

The second-longest-period variable with $P=114$ days, \#44277 is cool but not
sufficient to show TiO bands strong enough to be classified as an M-type star.
The {\sc atlas9} model fit is quite good and indicates a low temperature,
$T_{\rm eff}=3750$ K as well as a low metallicity, possibly [Fe/H]$<-2$. It is
with $B-V=1.84$ mag one of the very reddest stars in $\omega$\,Cen. This star
as well as \#44262, \#33062 and \#35250 have recently been found in {\it
Spitzer Space Telescope} images to have substantial amounts of circumstellar
dust (Boyer et al.\ 2007; McDonald et al., in preparation).

\subsubsection{Dredge-up}

%
%
\begin{figure}
\centerline{\psfig{figure=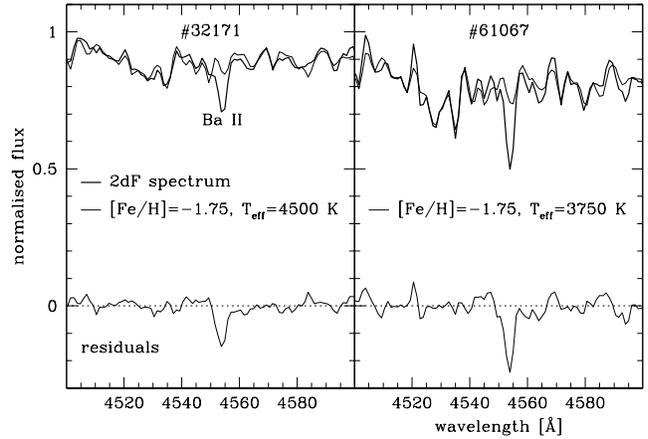,width=84mm}}
\caption[]{Two examples of barium enrichment.}
\end{figure}

An s-process element, barium has a strong line at 4554 \AA\ (Ba\,{\sc ii}). We
define the following index:
\begin{equation}
Ba4554 = -2.5 \log\left( \frac{2 F_{4553-4555}}{F_{4542-4544}+F_{4560-4562}}
\right),
\end{equation}
where the average flux levels, $F_{\lambda1-\lambda2}$ (between wavelengths
$\lambda1$ and $\lambda2$ in \AA), are measured in the normalised spectrum.
Many red giant stars show this line, and it becomes stronger in cooler stars.
A relatively small number of stars have much stronger barium lines than
expected from model atmospheres that do reproduce much of the remainder of the
spectrum (two examples are given in Fig.\ 22). These stars may thus be
enriched in s-process elements, possibly as a result of $3^{\rm rd}$ dredge-up
on the AGB. All M-type stars have exceptionally strong barium lines (Fig.\
20b).

The strength of the CH absorption extends significantly beyond the narrow
range in values of the cool M-type stars (Fig.\ 20c). Some of these are also
enriched in barium (Fig.\ 20d); none of the stars with CH weaker than that
found in M-type stars show any indication of barium enrichment. This clearly
suggests that the CH-strongest stars are not simply very cool, but must be
enriched in carbon. Alternatively, the M-type stars might be enriched in
oxygen, locking away more carbon atoms inside the CO molecule.

The CN bands in M-type stars are even weaker than the CH bands, compared to
other cool giants (Fig.\ 20e). The CN-strong stars may be enriched in
nitrogen, as well as carbon. There is no correlation whatever between barium
enrichment and strong CN bands (Fig.\ 21f), suggesting that the nitrogen
enrichment is not due to $3^{\rm rd}$ dredge-up.

%
%
\begin{figure}
\centerline{\psfig{figure=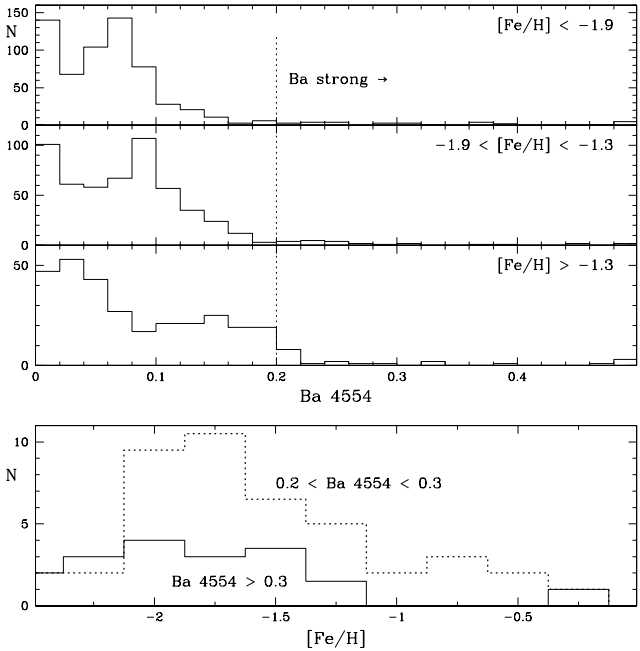,width=84mm}}
\caption[]{Barium line strength distributions for the metallicity intervals
[Fe/H]$<-1.9$, $-1.9<$[Fe/H]$<-1.3$ and [Fe/H]$>-1.3$ (top panels), and
metallicity distributions of the Ba-strong stars with $0.2<Ba4554<0.3$
(dotted) and $Ba4554>0.3$ (solid) in the bottom panel. Although metal-rich
stars have stronger barium lines, there is little evidence for a metallicity
bias of the Ba super-enhancement.}
\end{figure}

Stronger barium lines are found at higher metallicity (Fig.\ 23, top panels);
a distinctive peak in barium strength can be seen to progress from
$Ba4554\sim0.07$ in stars with [Fe/H]$<-1.9$, through $Ba4554\sim0.09$ in
stars with $-1.9<$[Fe/H]$<-1.3$, to $Ba4554\sim0.15$ in stars with
[Fe/H]$>-1.3$ (in stars with $Ba4554\,\lsim\,0.05$ the measurement may be
affected by blending lines of other elements). An increased barium abundance
amongst the metal-rich subpopulations in $\omega$\,Cen is well documented, and
likely due to enrichment by the ejecta from AGB stars over a period of star
formation lasting more than a Gyr (Smith et al.\ 2000). However, the Ba-strong
stars with $Ba4554>0.2$ show a very different picture (Fig.\ 23, bottom
panel). The metallicity distributions of these Ba-strong stars do not differ
very much from the overall cluster metallicity distribution (Fig.\ 10). We
will refer to this phenomenon as barium ``super-enhancement'', to distinguish
it from the metallicity-correlated chemical pre-enrichment seen at
$Ba4554<0.2$. Barium super-enhancement seems to be common amongst stars
belonging to the bulk of $\omega$\,Cen's population with [Fe/H]$\sim-1.8$, and
{\it not} a feature specific to a metal-rich subpopulation. A plausible
explanation would therefore be that barium super-enhancement is due to
dredge-up of s-process elements in the stars themselves.

%
%
\begin{figure}
\centerline{\psfig{figure=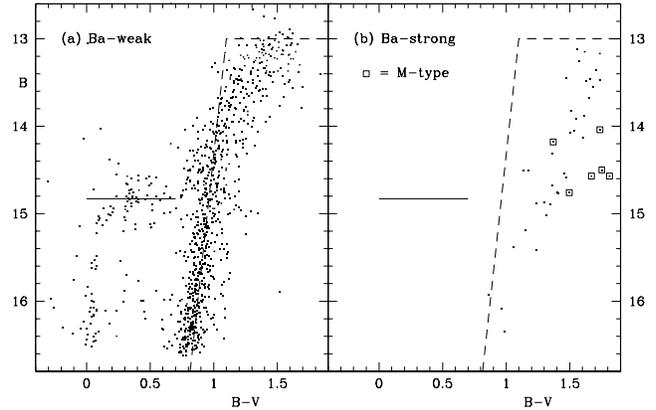,width=84mm}}
\caption[]{Colour-magnitude diagram of (a) Ba-weak stars and (b) Ba-strong
stars. The M-type stars are also indicated (squares).}
\end{figure}

The Ba-strong stars as defined in Fig.\ 20f are cool and luminous (Fig.\ 24);
if the barium is enriched due to $3^{\rm rd}$ dredge-up, then these stars
delineate the thermal-pulsing AGB. The reason why the Ba-strong sequence is a
little redder and fainter in the B-band than other, Ba-weak stars may be a
result of larger bolometric corrections to the optical photometry. The M-type
stars are also relatively faint in the B-band, probably for the same reason.
In fact, four of the six M-type stars are on the RGB-a sequence (Fig.\ 24).

\subsection{Carbon stars}

Apart from many stars which show the 4300 \AA\ CH band and the 3839 \AA\ CN
band, a few stars also show absorption by C$_2$ molecules around 4700 \AA\
(which form part of the Swan system). Whereas CH and CN are seen in cool,
oxygen-rich stars including the Sun, C$_2$ molecules only form when the
carbon-to-oxygen ratio is larger than unity. They have been called ``cool CH
stars'' in the past, but this is a misnomer as many cool CH stars do not show
C$_2$ bands. We prefer to refer to the stars showing C$_2$ bands as ``carbon
stars''. Four carbon stars were known in $\omega$\,Cen: ROA\,55 (Harding
1962), ROA\,70 (Dickens 1972; Wing \& Stock 1973), ROA\,279 (Bond 1975), and
ROA\,577 (Cowley \& Crampton 1985). Another carbon star, ROA\,153 has the
wrong proper motion (Woolley 1966) and radial velocity (Smith \& Wing 1973) to
be a cluster member.

%
%
\begin{figure}
\centerline{\psfig{figure=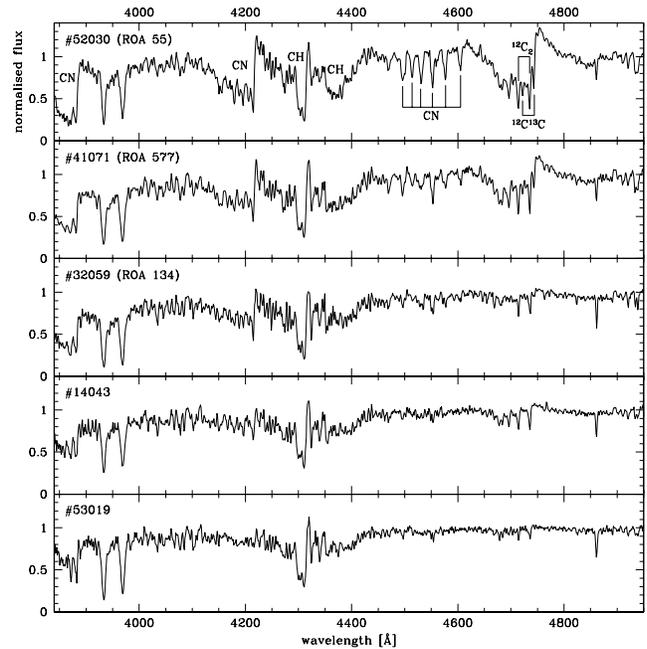,width=84mm}}
\caption[]{Spectra of the carbon stars detected in our survey. These are
defined as stars with absorption by C$_2$ molecules (most notably around 4700
\AA), not just CN and/or CH.}
\end{figure}

We obtained spectra of ROA\,55 (\#52030) and ROA\,577 (\#41071), and
discovered three more carbon stars: \#32059 (ROA\,134), \#14043 and \#53019
(Fig.\ 25). The measured stellar parameters were $T_{\rm eff}\sim4500$ K,
$\log(g)\sim1.5$ and [Fe/H]$\sim-2.5$ to $-2$, respectively, but these are
unreliable because the model does not reproduce the C$_2$ and strong CN bands.
The three new carbon stars would have been inconspicuous on the objective
prism plates that were used to survey $\omega$\,Cen in the past, and they
could only be found by a systematic spectroscopic survey. Especially the last
two stars only show moderate enhancement of the CN and CH bands. We may thus
anticipate that the total number of (weak) carbon stars in $\omega$\,Cen may
lie around $\sim30$.

The carbon stars are found along the entire length of the red giant branch
(Fig.\ 15). At least the fainter carbon stars are therefore more likely to
have had their carbon abundances enhanced through the process of mass transfer
from a close carbon star companion many Gyr ago. The three brightest carbon
stars show also the 4722+4744 \AA\ $^{12}$C$^{13}$C heavy-isotope version of
the 4714+4735 \AA\ C$_2$ lines; the $^{12}$C$^{13}$C:C$_2$ strength ratio is
very high in the brightest and reddest carbon star \#52030 (ROA\,55). This
suggests a $^{12}$C:$^{13}$C ratio $\ll10$, which clearly points at the
s-process in an AGB carbon star to have been responsible for the bulk of the
carbon found in this star. The Ba\,{\sc ii} 4554 \AA\ line is strongest in the
carbon stars with the strongest C$_2$ and CN bands, though, suggesting a link
between the carbon enrichment and an enhancement of the barium abundance.

\subsection{Post-AGB and UV-bright stars}

%
%
\begin{figure*}
\centerline{\psfig{figure=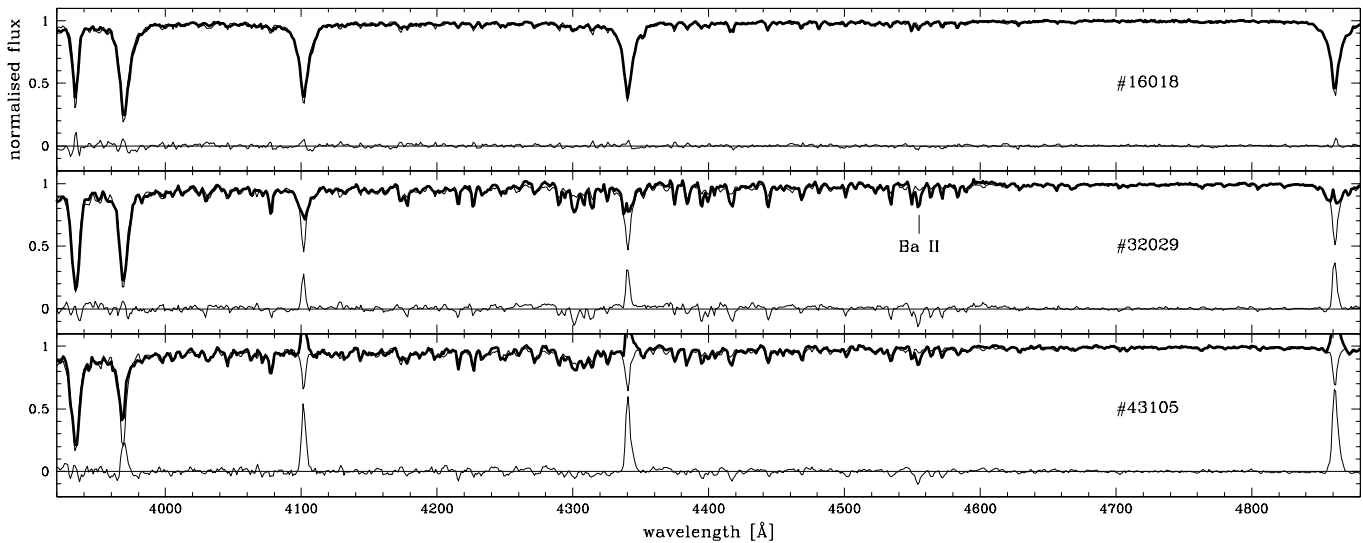,width=180mm}}
\caption[]{Spectra of post-AGB stars \#16018, \#32029 and \#43105 (bold
lines), together with the model fits (thin lines) and residuals.}
\end{figure*}

Stars brighter than the HB but not on the upper RGB or AGB are candidate
post-AGB stars (Fig.\ 12). For a handful the evidence for this is quite
convincing. Line emission is seen in \#43105 (V29) and \#32029 (Fig.\ 21),
indicative of circumstellar material reminiscent of a planetary nebula. The
line emission in \#32029 appears as line core inversions because of the lower
contrast against the brighter underlying stellar photospheric emission as
compared to \#43105. Very faint line emission might be present in the luminous
yellow supergiant \#16018 as well (Fig.\ 26), but it is at a level at which
the accuracy of the model fit cannot be guaranteed. In \#43105 and especially
\#32029 many absorption lines including the Ba\,{\sc ii} line at 4555 \AA\ are
stronger than in the model, which reproduces the calcium doublet and various
other lines rather well (Fig.\ 26). Indeed, Gonzalez \& Wallerstein (1994)
found CNO and s-process enhancements in the atmospheres of all these three
luminous UV-bright stars, confirming their post-AGB nature.

The fainter UV-bright stars do not show obvious line emission. They are all
confirmed proper motion and radial velocity members. Landsman et al.\ (1992)
argue that these stars are less luminous than one would expect post-AGB stars
to be, and they may instead have evolved of off the early-AGB (before the
onset of thermal pulses) or HB before climbing the AGB (AGB Manqu\'e). This is
consistent with the findings of Gonzalez \& Wallerstein (1994), who did not
detect any peculiarities in the chemical abundances of two of these stars,
\#32015 (ROA\,342) and V48. The absence of line emission may then be due to
weaker mass loss, or because of longer evolutionary timescales leading to the
dispersal of circumstellar material before it is ionised.

\section{Discussion}

\subsection{Late stages of evolution of metal-poor stars}

The late stages of stellar evolution determine the fate of a star and its
remnant (if any), as well as the return of nuclear processed material into the
ISM. Mass loss is a critical factor in this as it truncates stellar evolution.
The evolution of stars of nearly solar mass is particularly sensitive to the
time-dependence of the mass loss, as the mass loss during the RGB determines
how many thermal pulses on the AGB it will undergo --- if any (for a
comprehensive review of stellar evolution in globular clusters see Renzini \&
Fusi Pecci 1988).

\subsubsection{Horizontal Branch stars and RR\,Lyrae variables}

The core-helium burning phase is a valuable diagnostic intermediate phase in
post-main sequence evolution as the star's properties during this phase depend
critically on the core mass and mantle mass and composition. The core mass is
similar for all stars undergoing the helium flash at the tip of the RGB,
whilst the mantle mass depends on the integrated mass loss on the RGB. Stars
with thinner mantles become bluer HB stars, but a lower metallicity has a
similar effect as it reduces the opacity in the mantle rendering a hotter
photosphere. Indeed, $\omega$\,Cen displays a prominent metal-poor blue HB
(Fig.\ 12a, at $15<B<15.6$ mag) where there is a gap for metal-rich stars
(Fig.\ 12c). Metal-rich stars do become blue HB stars though (at $B\gsim15.5$
mag), possibly as a result of accelerated evolution if enhanced in helium (Lee
et al.\ 2005) or as a result of mass loss on the RGB. A smaller, bluer gap is
present for metal-poor stars (at $B\sim15.7$ mag). Differences in atmospheric
structure are expected as the ionization balance changes. This can be
accompanied by diffusion leading to surface metal enhancement, possibly
explaining the relatively high metallicity of the hottest (EHB) stars in our
sample, although metal-poor stars appear to populate the blue tail too (at
$B\gsim15.8$ mag).

Depending on the mass and mantle structure, some HB stars occupy the RR\,Lyrae
instability strip. The pulsation properties thus contain information related
to the RGB mass loss. For instance, Butler et al.\ (1978) suggest that some
metal-rich, later-spectral type RR\,Lyrae may be $\sim0.1$ M$_\odot$ less
massive, possibly due to heavier mass loss on the RGB. We do not find such
cooler metal-rich RR\,Lyrae, but we do find cool RR\,Lyrae stars with longer
pulsation periods ($P>1$ d) and lower surface gravities --- these may already
have evolved of off the zero-age HB and be moving up the AGB, and it is
possible that mass loss on the HB has further depleted their mantles.

Brighter HB stars could signal helium enrichment (e.g., Sweigart 1997). HB
stars about 0.2 mag brighter than the RR\,Lyrae locus are found amongst
relatively metal-rich stars (Fig.\ 12). On the other hand, Sollima et al.\
(2006) show that metal-intermediate RR\,Lyrae stars are fainter, and that they
are fully consistent with a normal helium abundance. Indeed, He-rich HB stars
are hotter and hence do not occupy the RR\,Lyrae instability strip (Lee et
al.\ 2005).

There is some variation amongst the warmer RR\,Lyrae, where the fraction of
overtone pulsators increases and where stars are found which have not so far
been recognised as RR\,Lyrae but whose surface gravities vary between repeat
measurements more than those of stable stars elsewhere in the HRD --- these
might be the low-amplitude pulsators predicted by Buchler \& Koll\'ath (2001).

\subsubsection{Asymptotic Giant Branch stars and dredge-up}

HB stars with sufficiently massive mantles will climb the AGB. The AGB blends
with the RGB, and in blue-visual colour-magnitude diagrams the tip-AGB stars
are no longer the brightest because they will be exceedingly cool and
therefore emit an increasing portion of their light at longer wavelengths. The
small group of M-type stars in our 2dF sample are redder but fainter than the
upper red giant branch, and most fall on the RGB-a anomalous red giant branch.
Our {\it Spitzer Space Telescope} infrared atlas, however, clearly shows that
they are the most luminous stars in the cluster (Boyer et al.\ 2007; McDonald
et al., in preparation). We thus identify the M-type stars as tip-AGB stars.
They appear to be nitrogen-poor, as the strength of the CN bands is very
modest at a given strength of the G-band (CH). The metallicities of these very
cool stars are difficult to determine reliably, but they do not appear to be
particularly high. The most luminous AGB stars in $\omega$\,Cen could have
their origin in the most primitive sub-population in the cluster. Evidence for
dredge-up of the products of nucleosynthesis in thermal pulses would
unambiguously confirm the AGB nature.

We have been able to isolate stars that are super-enriched in barium, above
the level resulting from pre-enrichment by a previous generation of AGB stars.
In the latter, comprising the majority of stars in which barium lines are
seen, the 4554 \AA\ barium line simply increases in concert with the CN and CH
band strength and with overall metallicity. This is expected as AGB stars with
masses $1.5<M<3$ M$_\odot$ produce a large amount of Ba and C (Smith et al.\
2000; Herwig 2005). The super-Ba-rich stars, though, do not show such
correlation (Figs.\ 20 \& 23), suggesting that low-mass, metal-poor stars may
be efficient in enriching their photospheres with s-process elements without
showing a marked carbon enrichment. The super-Ba-rich are generally the
coolest and brightest stars in the cluster, and include the M-type stars
(Fig.\ 24). We thus conclude that the super-Ba-rich stars in general, and the
M-type stars in particular, have undergone at least one thermal pulse, but
that this was not generally sufficient to raise the carbon-to-oxygen ratio
above unity. Perhaps a process is operating akin to the cool bottom processing
occurring in RGB stars, depleting carbon by burning it through the CNO cycle
at the bottom of the convective mantle when it dips into the hydrogen-burning
shell (Wasserburg, Boothroyd \& Sackmann 1995; Boothroyd \& Sackmann 1999).

Amongst the many stars with strong CN and CH bands we recover all three known,
and discover three new genuine carbon stars on the basis of the presence of
C$_2$ in their atmospheres. They are Ba-rich and $^{13}$C-rich, which suggests
dredge-up on the AGB. The faintest carbon star is unlikely to be on the AGB
though, and the $^{13}$C enhancement has also been explained by Origlia et
al.\ (2003) in terms of cool bottom processing and extra mixing on the RGB.
Whether or not the carbon stars that we find in $\omega$\,Cen today are the
product of self-enrichment or of mass transfer from an erstwhile more massive
carbon star companion, we do find carbon stars near the very tip of the AGB,
and it is likely that these will produce dust of a carbonaceous nature, which
will enter the ISM.

\subsubsection{Mass loss and post-AGB objects}

That said, the direct evidence in our medium-resolution spectra for mass loss
is meager. Shocks due to strong stellar pulsation are found in two M-type AGB
stars in the form of hydrogen line emission, but echelle spectra are needed to
detect outflow motion. An important question is whether the well-known
paradigm of pulsation-initiated, dust-assisted mass loss necessarily comprises
most of the mass shed by metal-poor low-mass stars. Such stars spend most of
their time being too warm for pulsations to develop and for an extended
molecular atmosphere and circumstellar dust to form. Yet these stars are
likely to have a chromosphere, which might provide an alternative means for
driving mass loss (Schr\"oder \& Cuntz 2005; McDonald \& van Loon, in
preparation). Our spectra do show hints of chromospheric emission in the
Ca\,H+K doublet along much of the red giant branch, but this diagnostic is too
sensitive to other effects to be reliably applied to our medium-resolution
spectra.

We identified post-AGB stars in $\omega$\,Cen on the basis of their blue
colours, high luminosities and in a few cases hydrogen line emission (besides
proper motion and radial velocity membership). Like the M-type AGB stars they
appear to be metal-poor (Fig.\ 12). We speculate that metal-poor stars might
suffer less mass loss on the RGB and hence climb higher on the AGB, ultimately
becoming post-AGB stars. This is consistent with the copious dust production
and presence of a carbon-rich planetary nebula in the extremely metal-poor
globular cluster M\,15 (Boyer et al.\ 2006). Metal-rich clusters would be
predicted to host more post-early-AGB and post-HB stars, as stronger mass loss
on the RGB would have reduced their mantles enough to prevent them from
reaching the thermal-pulsing part of the AGB. This is corroborated by our
data, which suggest that the less luminous UV-bright stars are relatively
metal-rich. On the other hand, however, metal-rich clusters, with
[Fe/H]$\simeq-0.7$, are known to host more luminous AGB stars undergoing
long-period variability which suggests that they have larger core masses
(Frogel \& Elias 1988). Either these clusters are younger by several Gyr, or
mass loss on the RGB was not stronger than for clusters with [Fe/H]$\sim-2$ to
$-1.5$ (cf.\ Renzini \& Fusi Pecci 1988).

\subsection{Cluster formation and evolution}

The details of the formation and evolution of $\omega$\,Cen are imprinted in
the chemical and dynamical properties of its multiple populations. We briefly
discuss the results from our analysis in this context.

\subsubsection{The nature of the multiple stellar populations}

The isolation and characterization of stellar populations of different age and
composition is hampered by the high level of degeneracy between metallicity
and temperature in photometry and low to medium-resolution spectroscopy. Below
$\sim4000$ K much of the atomic and molecular opacity increases both with
higher abundance and lower temperature. This could explain why our
spectroscopic analysis indicates that the cool RGB-a stars may have ``normal''
abundances, [Fe/H]$\sim-1.7$, where other published work indicates that the
RGB-a is considerably more metal-rich. Such discrepancies may also arise from
spectroscopic missions such as {\it GAIA}, so it is important to be aware of
the ways in which a different methodology could yield different answers.

Metal-rich stars tend to be identified through calcium/colour methods not by
means of a full spectral synthesis. We do not constrain the solutions by using
priors, but obtain the best fit to the observed spectrum by varying T$_{\rm
eff}$, $\log({\rm g})$ {\it and} [Fe/H]. Neither Origlia et al.\ (2003),
Sollima et al.\ (2005), Stanford et al.\ (2006a) or Villanova et al.\ (2007)
fit $T_{\rm eff}$ to the spectra but instead use photometry to constrain it
(Villanova et al.\ also constrain $\log({\rm g})$ in this way). Their results
are not always consistent, for instance Villanova et al.\ (2007) find
[Fe/H]$=-1.1$ for the subgiants that evolve into the RGB-a, which is 0.5 dex
lower than Pancino et al.\ (2002) find. The latter authors do estimate T$_{\rm
eff}$ directly from their spectra by imposing excitation equilibrium on a set
of iron lines. They analysed three metal-rich and three
intermediate-metallicity stars, and find temperatures around 4000 K, i.e.\ not
extremely cool. Our own analysis suggests that what look like RGB-a stars may
in fact be cool but metal-poor. As our data do detect a metal-rich
sub-population across the upper HRD, we tentatively conclude that our low
[Fe/H] values for the RGB-a stars may point at a problem inherent to the
spectroscopic technique (or sample selection).

%
%
\begin{table}
\caption[]{Comparison between the solutions for T$_{\rm eff}$ and [Fe/H] with
the original procedure (subscripts ``0'') and when fixing the value for
T$_{\rm eff}$ to that closest to the value derived from the Alonso et al.\
(1999) colour-T$_{\rm eff}$ relation (subscripts ``a''), for a selection of
stars with colours like RGB-a stars: $14.4<B<15.1$ and $B-V>1.3$.}
\begin{tabular}{llllll}
\hline\hline
LEID & $B-V$ & $T_{\rm eff, a}$ (K\rlap{)} & [Fe/H]$_{\rm a}$ &
$T_{\rm eff, 0}$ (K\rlap{)} & [Fe/H]$_0$ \\
\hline
32149 & 1.302 & 4169 & $-1.50$ & 4000 & $-1.75$  \\
18046 & 1.319 & 4147 & $-1.25$ & 4000 & $-1.50$  \\
69027 & 1.399 & 4048 & $-1.50$ & 4000 & $-1.375$ \\
23031 & 1.347 & 4112 & $-1.25$ & 4000 & $-1.25$  \\
77010 & 1.351 & 4107 & $-2.00$ & 5000 & $-1.00$  \\
48321 & 1.360 & 4096 & $-2.00$ & 3750 & $-2.25$  \\
54022 & 1.412 & 4033 & $-1.50$ & 4000 & $-1.50$  \\
48323 & 1.461 & 3975 & $-1.75$ & 3750 & $-2.00$  \\
60058 & 1.475 & 3958 & $-1.50$ & 4000 & $-1.50$  \\
44484 & 1.497 & 3933 & $-2.25$ & 3500 & $-2.50$  \\
35094 & 1.670 & 3746 & $-1.75$ & 3500 & $-2.00$  \\
35250 & 1.752 & 3663 & $-1.75$ & 3500 & $-2.25$  \\
33062 & 1.813 & 3604 & $-2.25$ & 3500 & $-2.00$  \\
\hline
\end{tabular}
\end{table}

To investigate the extent to which the freedom in our method may produce less
realistic solutions than methods in which some of the parameters are set {\it
a priori}, we have run our model fitting procedure on the 13 RGB-a candidate
stars with $14.4<B<15.1$ and $B-V>1.3$ but constraining the solution for
T$_{\rm eff}$ to that closest to the value predicted by the Alonso et al.\
(1999) colour-T$_{\rm eff}$ relation. The results are listed in Table 4 (with
subscripts ``a'') and compared to the solutions obtained with the unrestrained
procedure (with subscripts ``0''). The differences are generally small, only
one step in temperature and/or metallicity. This is because the original
procedure yields temperatures that are already in fair agreement with the
colour-T$_{\rm eff}$ relations. Although in six cases the value for [Fe/H] has
increased by constraining the value for T$_{\rm eff}$, in three cases it has
actually decreased. The formerly most metal-rich star in this selection,
\#77010 has become one of the most metal-poor stars. This is the only star for
which the value for the gravity changed: from $\log(g)=2$ to become 0. We
conclude that setting constraints on the available parameterspace informed by
photometry has a systematic but not very dramatic effect on the metallicity
determinations, and does not necessarily result in higher metallicities for
stars in the RGB-a portion of the (B, B--V) colour-magnitude diagram.

Irrespectively, our selection of RGB-a stars is not uniformly strong in CN. In
fact, some M-type stars are found on the RGB-a. These are cool, which would
have made CN bands strong, but clearly oxygen-rich, which agrees with the
observed weak CN bands. Pancino et al.\ (2002) and Origlia et al.\ (2003)
indeed find that RGB-a stars have solar [$\alpha$/Fe], and suggest the delayed
enrichment with iron by supernovae of Type Ia. These stars must then also be
pre-enriched in s-process elements by a previous generation of AGB stars. This
scenario is confirmed by the increasing strength of the barium line with
increasing metallicity (Fig.\ 23), which would be more consistent if the RGB-a
stars were metal-rich. The CN-rich stars are generally metal-rich as well as
nitrogen-enriched --- compared to M-type stars and CN-weak stars in general.
The elevated levels of nitrogen must be due to pre-enrichment in these stars.
The helium-rich stars are probably related to this episode of formation from
gas enriched by massive stars. Different evidences suggesting a large helium
enhancement of RGB-a stars have been put forward (Norris et al.\ 1996; Sollima
et al.\ 2005; Lee et al.\ 2005; Villanova et al.\ 2007).

We conclude that care has to be exercised in studying the RGB-a, as extreme
AGB stars may mimic RGB-a optical colours and magnitudes. In combination with
the difficulty in modelling accurately the stellar atmospheres and spectra of
cool stars and measuring the stellar properties from low-resolution spectra,
this explains our general finding that RGB-a stars may be metal-poor whilst
all other evidence suggests that (true) RGB-a stars are metal-rich.

\subsubsection{Gas retention and accretion}

Our data independently confirm some of the internal kinematic properties of
$\omega$\,Cen as reported in the literature, in particular the rotation of its
core (Merritt et al.\ 1997; van Leeuwen \& Le Poole 2002; Reijns et al.\
2006). We do not find evidence for different kinematics of the various
sub-populations, nor for a different spatial distribution. Sollima et al.\
(2007) find that the blue (intermediate metal-rich) main sequence stars are
more concentrated towards the cluster centre, confirming an earlier result
obtained by Norris et al.\ (1997) and for the metal-rich stars by Pancino et
al.\ (2003). Norris et al.\ (1997) and Ferraro et al.\ (2002) also suggested a
difference in kinematics between the metal-poor and metal-rich populations,
but van de Ven et al.\ (2006) and Pancino et al.\ (2007), as well as our own
data, show that there is no evidence for such difference thus removing the
strongest evidence for a merger scenario to explain the multiple populations
in $\omega$\,Cen. The more likely scenario for producing the multiple
populations in $\omega$\,Cen thus remains the prolonged or episodic star
formation as a result of gas retention or accretion in the immediate
proto-cluster environment, possibly as part of a once larger system (cf.\
Bekki 2006).

We obtained a crude map of the diffuse ionized medium in the foreground of
$\omega$\,Cen, demonstrating the potential of using the many blue HB stars in
metal-poor globular clusters for probing the ISM. We also tentatively detect,
in both the H and K components of the Ca\,{\sc ii} doublet, what appears to be
a high-velocity cloud moving towards $\omega$\,Cen at a (projected) velocity
close to 200 km s$^{-1}$ with respect to the systemic motion of the cluster.
This is faster than the expected free-fall velocity, which would be $\sim42$
km s$^{-1}$ to reach the inner half-mass radius of $\sim6$ pc if the total
mass is $\sim2.5\times10^6$ M$_\odot$ (van de Ven et al.\ 2006). The cluster
is therefore likely to pass through it and leave it behind. Slower encounters
could lead to capture of interstellar clouds, and we speculate that accretion
events could in the distant past have led to the formation of the chemically
distinct sub-populations that we see today.

\section{Summary of conclusions}

We obtained medium-resolution optical spectra for $>1500$ proper motion
members of the Galactic globular cluster $\omega$\,Centauri, making a
particular effort to sample the entire upper part of the optical
colour-magnitude diagram (not just where most stars are). We measured the
radial velocity, effective temperature, metallicity and gravity in an
automated fashion by fitting synthetic spectra based on model atmospheres. We
present a catalogue which contains these data as well as the measured line
strengths of Ca\,{\sc ii} H+K, Ba 4554 \AA, and CN, CH and TiO.

The radial velocities confirm membership for nearly all stars, and display the
known rotation of the cluster core. The metallicity distribution peaks around
[Fe/H]$\sim-1.8$, with a secondary peak around [Fe/H]$\sim-0.8$.

The RR\,Lyrae stars pulsating in the $1^{\rm st}$ overtone are found to be
warmer on average than those pulsating in the fundamental mode. Four stars
pulsating with periods $P\,\gsim\,1$ d have relatively low gravities, either
because they have evolved towards the AGB and/or their masses are reduced as a
result of mass loss. We tentatively identify a sample of low-amplitude
variables in the blue part of the RR\,Lyrae instability strip on the basis of
their relatively large variations in gravity between multiple measurements.
For one RR\,Lyrae star we find blue-shifted hydrogen Balmer line emission.

Measurements of metallicities and temperatures purely on the basis of our
spectroscopy do not always agree with the more commonly used methods in which
the temperature is constrained with photometry; e.g., the anomalous RGB
appears metal-poor in our data, contrary to what is generally accepted. This
is likely the result of the inherent difficulty in modelling the atmospheres
and spectra of cool stars, and the intrusion into the RGB-a optical colours
and magnitudes regime by extreme AGB stars.

We identify several new cluster carbon stars, and find evidence for thermal
pulses enriching M-type AGB stars with barium. This super-enhancement is
distinct from the pre-enrichment in s-process elements by a previous
generation of more massive AGB stars, which our measurements also confirm to
have taken place.

We also identify several post-AGB stars and other UV-bright stars; their and
the AGB star properties can be interpreted as mass loss on the RGB being
slightly less efficient around [Fe/H]$\sim-2$ than around [Fe/H]$\sim-1$.
Evidence for mass return into the ISM is present in the form of hydrogen line
emission from post-AGB stars, and we postulate that some of the brightest
carbon stars might eventually inject carbon-rich dust into the ISM.

Blue HB stars are used to probe the intervening ionized ISM in the Ca\,{\sc
ii} H+K lines, revealing possible interaction between the cluster and
surrounding diffuse matter. Higher-resolution spectra of more blue HB stars
are required to confirm these results.

\section*{Acknowledgments}

We wish to thank Terry Bridges for performing the 2dF observations in his
capacity as erstwhile 2dF fellow, and for his advice during the fibre
configuration and data reduction stages. We are grateful to the referee whose
insightful and constructive review helped clarify certain points in the
manuscript. AS acknowledges the award of a Nuffield Foundation Undergraduate
Research Bursary to perform part of this research. IMcD acknowledges an STFC
studentship to fund his research on mass loss in globular clusters. This
project was initiated while JvL was a PDRA at the IoA Cambridge, the
hospitality of which is much appreciated.


\appendix

\section{Minimising differential Taylor series expansions in practice}

The observed spectrum, $f(x)$ is presumed to be described approximately as a
Taylor series expanded to second order around each spectral point $a$,
\begin{equation}
f(x) \simeq f(a) + (x-a) \left. \frac{{\rm d}f}{{\rm d}x} \right|_{x=a} +
\frac{(x-a)^2}{2!} \left. \frac{{\rm d}^2f}{{\rm d}x^2} \right|_{x=a},
\end{equation}
and the model spectrum, $g$, is written in identical fashion. Rather than
striving to obtain $f(a)\equiv g(a)$, we search for the model which yields
$f(x)\equiv g(x)$, in every point of the spectrum. Thus, with $N$ spectral
points we minimise the statistic
\begin{equation}
\chi^2 \equiv \sum_{i=1}^{N} \left(f_i(x)-g_i(x)\right)^2.
\end{equation}
In practice, we choose to evaluate the series expansion at a positive
deviation by one spectral point, i.e.\ $(x-a_i)\equiv1$. The $0^{\rm
th}$-order term is simply the difference between the model and observed
spectrum at $x=a_i$,
\begin{equation}
\Delta_i = f_i(a_i)-g_i(a_i),
\end{equation}
the $1^{\rm st}$-order term corresponds to the slope around $x=a_i$,
\begin{equation}
\Delta^\prime_i \simeq \frac{\Delta_{i+1}-\Delta_{i-1}}{2},
\end{equation}
whilst the $2^{\rm nd}$-order term is estimated from the difference in slope
at either side of $x=a_i$, i.e.\ in points $a_{i+1}$ and $a_{i-1}$,
\begin{equation}
\Delta^{\prime\prime}_i \sim \frac{\Delta_{i+2}-\Delta_i}{2} -
\frac{\Delta_i-\Delta_{i-2}}{2}.
\end{equation}
Combination of Eqs.\ (A3-5) yields
\begin{equation}
\chi^2 \sim \sum_{i=1}^{N} \left( \frac{\Delta_{i+2} +\Delta_{i+1}
-\Delta_{i-1} +\Delta_{i-2}}{2} \right)^2.
\end{equation}
If the line core has a different slope from the model then it will increase
the value of $\chi^2$; if it is simply shallower then it will not contribute
to $\chi^2$, but such discrepancies will, of course, contribute to $\chi^2$ in
adjacent points. A conspiracy could arise if $\Delta_{i+2}$ and $\Delta_{i-2}$
cancel each other and $\Delta_{i+1}$ and $\Delta_{i-1}$ are identical but not
zero. Clearly the spectral shape around that point would be quite different in
the model and observed spectra, but although this might occur in individual
spectral points the overall spectrum would be sufficiently different that it
is highly unlikely to minimise $\chi^2$ over the full spectrum.

Monte Carlo simulations were performed in which noise was added to a variety
of spectra produced from the {\sc atlas9} models. The method recovered the
input, never deviating by more than one step as long as the S/N ratio exceeded
$\sim10$. The model spectra are highly idealised versions, though, and in
reality stellar spectra deviate for many reasons. The traditional $\chi^2$
method on a pixel-by-pixel basis is susceptible to single, strong spectral
features such as the CN bands or Ca\,H+K lines. These were sometimes noticed
to drive the solution into an extreme corner of parameterspace whereas our
modified $\chi^2$ method would find a more ``expected'' solution (e.g.,
[Fe/H]$\sim-1.75$, or low gravity in case of a bright cluster red giant or
high gravity in case of an extreme HB star, et cetera). Our method thus seems
more robust than the traditional $\chi^2$ minimisation, as it takes account of
correlated behaviour due to spectral structure.

%
%
\begin{figure}
\centerline{
\hbox{
\psfig{figure=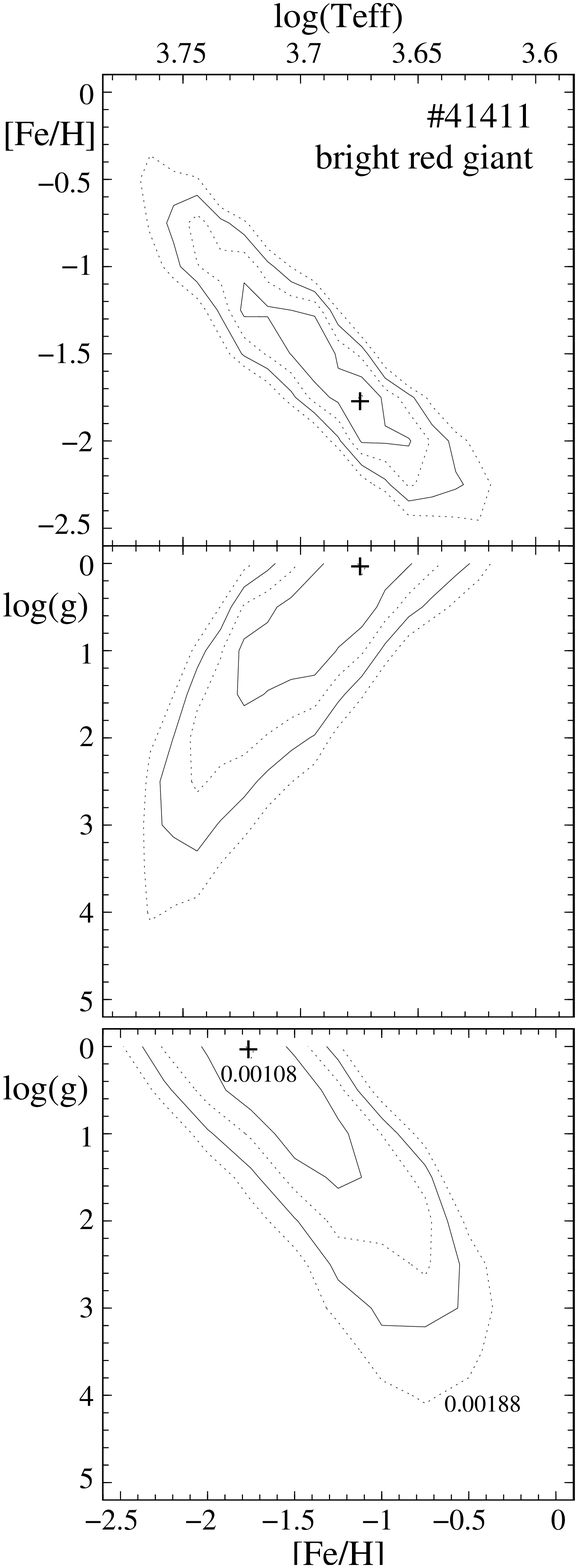,width=30.75mm}
\psfig{figure=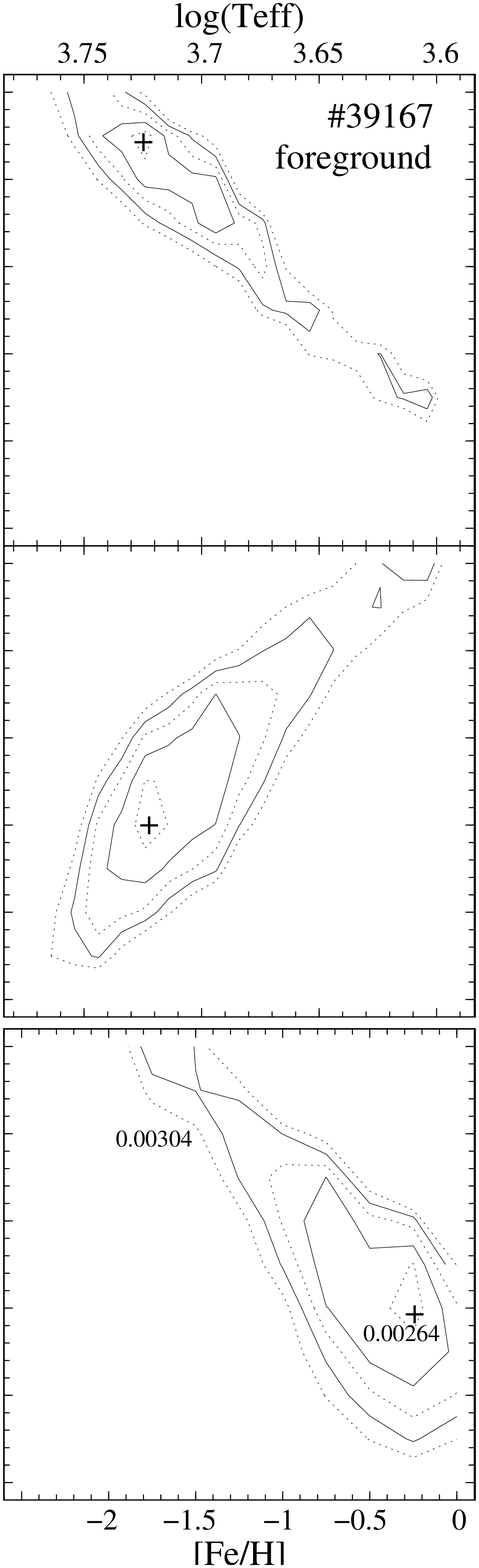,width=25.5mm}
\psfig{figure=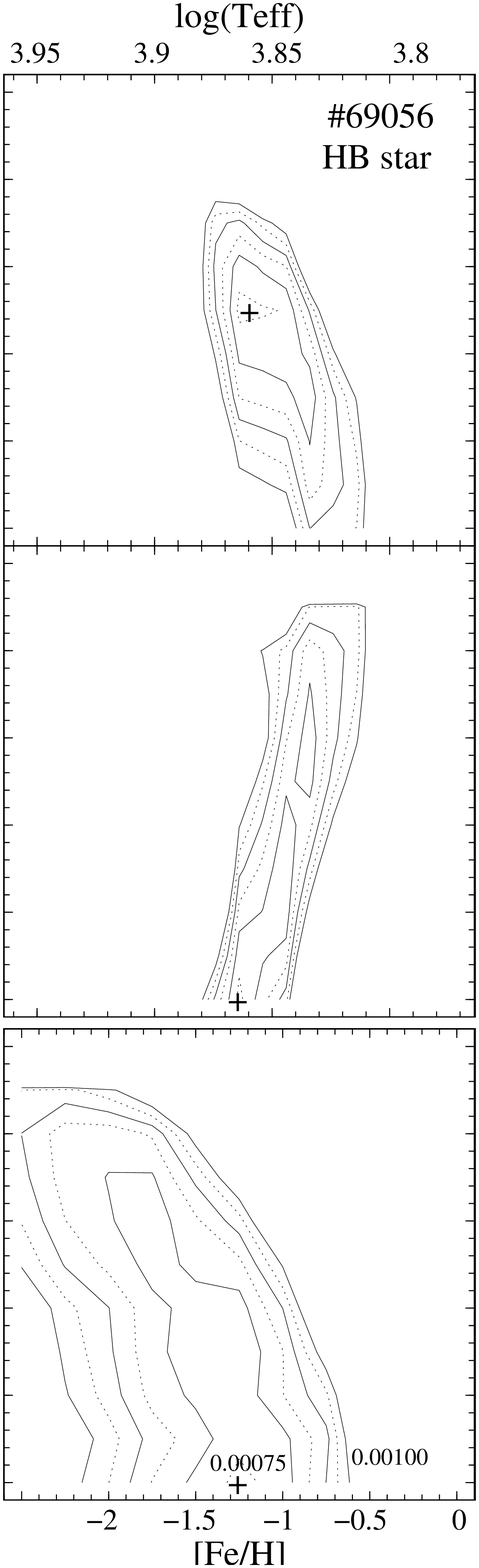,width=25.5mm}
}}
\caption[]{Maps of $\chi^2$, for the cluster red giant \#41411 (left),
foreground star \#39167 (middle) and cluster HB star \#69056 (right). Results
are displayed for the best-matched velocity and combinations of two parameters
$\in(T_{\rm eff}, [{\rm Fe/H}], \log(g))$ always wherever the third parameter
reaches the minimum value for $\chi^2$. Lowest and highest contours are
labelled; ``+'' marks minimum.}
\end{figure}

To illustrate the contrast of $\chi^2$ in parameterspace, $\chi^2$ maps are
shown in Fig.\ A1 for three cases, always for the best-matched velocity. In
each map, for each pair of values of the two parameters on the axes the
minimum value of $\chi^2$ is plotted, i.e.\ it is not a simple cross-section
through the $(T_{\rm eff}, [{\rm Fe/H}], \log(g))$ parameterspace but a
projection of where $\chi^2$ is minimal. This more accurately maps the space
where solutions tend to migrate, but it means that the corresponding value of
the third parameter varies across each map. Cooler solutions tend to be
accompanied by lower metallicity to offset the diminished line opacity. The
method easily distinguishes between a metal-poor cluster red giant (Fig.\ A1,
left panels) and a metal-rich foreground star (Fig.\ A1, middle panels). The
gravity for HB stars (Fig.\ A1, right panels) is less well constrained though
this hardly affects the solution for $T_{\rm eff}$ and to a lesser extent
[Fe/H].

\label{lastpage}

\end{document}